\documentclass[%
 reprint,longbibliography,preprintnumbers,
nofootinbib,
 amsmath,amssymb,
 aps,superscriptaddress,prd
]{revtex4-2}

\usepackage{graphicx}
\usepackage{dcolumn}
\usepackage{bm}

\usepackage{overpic}
\usepackage[english]{babel}
\usepackage{graphicx}
\usepackage{epsfig}
\usepackage{amssymb}
\usepackage{amsmath}
\usepackage[plainpages=false,pagebackref=false,pdftex]{hyperref}
\usepackage{ulem}
\usepackage[T1]{fontenc} 
\usepackage[utf8]{inputenc}
\usepackage{color}

\usepackage[titletoc]{appendix}
\usepackage{caption}
\usepackage{subcaption}  
\usepackage{float}

\usepackage{booktabs}

\begin{document}
\title{Scaling functions in the soft-wall AdS/QCD models}

\author{Zhongzheng Zhang}
\author{Danning Li}
 \vspace{1cm}
\affiliation{Department of Physics and Siyuan Laboratory, Jinan University, Guangzhou 510632, China}

\author{Lang Yu}
\affiliation{College of Physics, Jilin University, Changchun 130012, P.R. China}

\author{Zhibin Li}
 \vspace{1cm}
\affiliation{ Institute for Astrophysics, School of Physics, Zhengzhou University, Zhengzhou 450001, China}

\author{Xinyang Wang}
\affiliation{Fundamental physics center, School of Mechanics and Physics, Anhui University of Science and Technology, Huainan, Anhui 232001, China}

\begin{abstract}
We investigate the static scaling behavior of the chiral condensate near the two-flavor critical point within the framework of the soft-wall AdS/QCD. The scaling functions are extracted from the chiral order parameters and are found to precisely match those obtained through mean-field calculations. Additionally, it is also checked that the scaling functions are independent of the specific construction of the holographic model. Furthermore, we develop the formalism for calculating the chiral susceptibility and demonstrate that the pseudo-critical temperatures obey the scaling law for moderate quark masses. It is shown that the temperature scaling could be comparable with those obtained from Dyson-Schwinger equations and lattice simulations. These findings could help improve the effectiveness of the soft-wall AdS/QCD.
\end{abstract}

\keywords{ Soft-wall AdS/QCD, scaling function, critical temperature}

\maketitle

\section{Introduction}\label{secion:introduction}

Empirical and lattice-QCD studies indicate that, at non-zero temperature and baryon chemical potential, Quantum Chromodynamics (QCD) undergoes two intertwined but conceptually distinct phase transitions: deconfinement, whereby color degrees of freedom become manifest over hadronic scales, and chiral symmetry restoration, characterized by the melting of the quark condensate. Clarifying the interplay between these two phenomena is essential for interpreting experimental observables in heavy-ion collisions, elucidating the equation of state of dense matter in compact stars, and constraining scenarios for the evolution of the early Universe \cite{Braun-Munzinger:2003pwq,Shuryak:2014zxa,Rischke:2003mt,Huang:2023ogw}.

For physical quark masses, lattice-QCD studies establish that QCD matter undergoes a smooth analytic crossover at vanishing baryon chemical potential~\cite{Pisarski:1983ms,Borsanyi:2010bp,Cheng:2006qk,Ding:2015ona}. In the two-flavor chiral limit, by contrast, the transition becomes a true phase transition whose order---second or first---depends on the relative hierarchy between chiral-symmetry restoration and the effective restoration of the axial $U(1)$ anomaly \cite{Pisarski:1983ms}. Besides its theoretical significance, mapping the QCD phase diagram in the quark-mass plane is essential for locating the critical end-point (CEP), a primary objective of present heavy-ion-collision experiments; in particular, the chiral-limit transition temperature provides an upper bound for possible CEP temperatures at non-zero baryon chemical potential \cite{Bai:2020ufa}. Since lattice simulations cannot be carried out directly in the exact chiral limit, analyses that exploit universal scaling functions are indispensable for extrapolating physical results to that limit.

Recent lattice QCD results have revealed the anticipated  $O(N)$  scaling behavior~\cite{Kogut:2006gt,Ejiri:2009ac,Bazavov:2011nk} of the chiral crossover. The connection based on the universal scaling equation is restricted to a finite domain, referred to as the scaling window. The width of this window is still under debate. According to lattice QCD results~\cite{Burger:2011zc, Kotov:2021rah}, the scaling region is located at relatively large values of the pion mass. Ref.~\cite{Kotov:2021hri}, for example, explores the conformal window across different flavor numbers and suggests that, within the 3D $O(4)$ universality class, the physical pion mass may lie close to the critical point, with scaling behavior potentially extending into the high-temperature regime up to 300~MeV. In contrast, another lattice QCD study~\cite{HotQCD:2019xnw} indicates that the scaling window may also extend below the physical quark mass and, based on the scaling function, provides an estimate of the transition temperature in the chiral limit. Consistently, alternative approaches such as Dyson-Schwinger equations (DSE)~\cite{Bai:2020ufa,Gao:2021vsf,Bernhardt:2023hpr} and the functional renormalization group (FRG)~\cite{Braun:2010vd,Braun:2023qak} also demonstrate the presence of scaling behavior for quark masses below the physical point.	

In the low-energy regime where QCD phase transitions take place, the strong coupling strength renders conventional perturbative techniques inadequate. Thus, it is essential to develop non-perturbative methods to explore the underlying physics. Among such methods, lattice QCD---formulated from first principles---has been recognized as a reliable tool. However, at large baryon densities, lattice QCD suffers from the notorious sign problem~\cite{Fodor:2001au}, which has yet to be satisfactorily resolved. This limitation has motivated the search for alternative non-perturbative approaches. 

In the 1990s, 't Hooft proposed a preliminary form of the holographic principle~\cite{tHooft:1993dmi}, which was subsequently integrated with string theory by Susskind~\cite{Susskind:1994vu}. The AdS/CFT correspondence, first conjectured by Maldacena~\cite{Maldacena:1998zhr}, has shed new light on solving strongly coupled problems in gauge field theories.  In the study of QCD matter, a celebrated achievement of holographic approaches is the derivation of the universal lower bound  $\eta/s=1/(4\pi)$ for the shear viscosity to entropy density ratio, providing key insight into the nearly perfect-fluid behavior of the quark–gluon plasma from a theoretical perspective ~\cite{Policastro:2001yc,Buchel:2003tz,Kovtun:2004de}. Employing a bottom-up strategy that prioritizes phenomenology over top-down string-theoretic origins, researchers have constructed a variety of holographic models---including light-front holographic QCD \cite{Brodsky:15lfhQCD}, V-QCD~\cite{Gursoy:2017wzz}, and Einstein–Dilaton–Maxwell setups~\cite{Gubser:2008yx,Gursoy:2008bu,Li:2011hp,Finazzo:2014cna,Zollner:2018uep,Zhao:2022uxc,Chen:2024ckb}---to capture hadron spectra, thermodynamics, transport coefficients and the QCD phase transition with remarkable success. From those phenomenological studies, it appears that the EMD model provides an adequate description of gluodynamics.

 It has been shown that, spontaneous chiral symmetry breaking, as another key low-energy feature of QCD, can be effectively described within the hard-wall and soft-wall AdS/QCD models \cite{Erlich:2005qh,Karch:2006pv}. A finite chiral condensate is dynamically generated, lifting the degeneracy between chiral partners in the hadron spectrum \cite{Gherghetta:2009ac,Li:2013oda}. As the temperature increases, this condensate is progressively suppressed; once it vanishes, chiral symmetry is restored and the chiral phase transition takes place~\cite{Chelabi:2015gpc,Chelabi:2015cwn}. 

The principal advantage of this model is that, by elevating the four-dimensional (4D) global symmetry to a five-dimensional (5D) gauge symmetry, it naturally incorporates the conserved currents and the $\bar{q}q$ bilinear operator. This makes it straightforward to explore scenarios with arbitrary quark masses and finite charge densities. Studies have extended from simple two-flavor systems to  $N_f = 2+1$~\cite{Li:2016smq,Bartz:2017jku,Fang:2016nfj}, and more recently to  $N_f = 2+1+1$~\cite{Ahmed:2024rbj}. The phase diagram in the quark mass plane ($m_{u/d}-m_s$) can be summarized in a holographic version \cite{Chen:2018msc} of the so-called `Columbia plot', which qualitatively agrees well with that obtained by combining the lattice simulations and other effective studies \cite{Brown:1990ev,Ding:2015ona}. By introducing the baryon and isospin chemical potentials $\mu_B$ and $\mu_I$ through the corresponding conserved currents, one can probe regimes of high baryon and isospin densities where recent studies predict the appearance of additional phases--- e.g., the quarkyonic and pion-condensed phases \cite{Chen:2024cxh,Lv:2018wfq,Nishihara:2014nva,Chen:2019rez}. A magnetic field can be introduced by constructing the electromagnetic current from the baryon and isospin currents; inverse magnetic catalysis then emerges naturally, without fine-tuning of the model settings \cite{Li:2016gfn,Rodrigues:2018pep}. Furthermore, as derived in \cite{Colangelo:2012ipa}, the model can be reduced to a four-dimensional chiral perturbation theory. The matching of the near $T_c$ behavior of its Goldstone modes \cite{Cao:2020ryx,Cao:2021tcr,Cao:2022csq} with those of the four-dimensional chiral perturbation theory \cite{Son:2001ff} provides explicit evidence for this equivalence. Overall, the soft-wall model offers an excellent starting point for exploring chiral dynamics in QCD, serving in effect as a five-dimensional analogue of chiral perturbation theory.

Although the above analyses have thoroughly validated the soft-wall model’s consistency and effectiveness in describing low-energy QCD, its universality has yet to be fully tested.  While the critical exponents have been proved to be of mean field level \cite{Chen:2018msc}, with $\beta=1/2, \delta=3$ in the two-flavor limit, the holographic investigation of mass scaling behavior of the pseudo-critical temperature, as well as of the scaling function itself, remains largely undeveloped. Therefore, in this work, we aim to investigate the scaling behavior of the chiral phase transition temperature $T_c$ with respect to different quark and pion masses. We will calculate the scaling function and test the scaling window in the holographic model, which may provide further constraints for building a realistic holographic QCD model. In this work, we will focus on the two-flavor limit for simplicity.

The organization of this paper is as follows. In Sec.~\ref{section:model}, we briefly review the soft-wall model and chiral phase transition in this model. In Sec.~\ref{secion:scaling funciton}, we numerically extract the scaling functions and test the universal relations of those functions as a check of the self-consistency of the soft-wall AdS/QCD models. In Sec.~\ref{section:tc scaling}, we give analyses on the scaling behavior of the pseudo-critical temperature. An additional constraint is given for constructing a more realistic holographic QCD model for chiral dynamics. Finally, a brief summary will be given in Sec.~\ref{CONCLUSION AND DISCUSSION}.

\section{The soft-wall AdS/QCD model}\label{section:model}

In this section, we provide a brief overview of the soft-wall model and introduce several modifications to explore its universal properties. As discussed above,  the action of the soft-wall model is constructed by promoting the 4D global chiral symmetry $SU(2)_L \otimes SU(2)_R$ to 5D gauge symmetry. Since we focus on the two-flavor limit, i.e., considering only the u and d quarks, the action of the soft-wall model is given as follows:

\begin{align}
S_{5\mathrm{D}} = \int d^5x  \sqrt{g} e^{-\Phi} \mathrm{Tr}\Big\{ 
& |D_M X|^2 - V(|X|) \notag \\
& - \frac{1}{4g_5^2}(F_L^2 + F_R^2) 
\Big\}.
\end{align}

In the above action, $M$ represents spacetime indices, $g$ is the determinant of the metric  $g_{MN}$, $X$ is an $N_f \times N_f$ matrix-valued scalar field, and $\Phi$ is the dilaton field. As in previous studies, we set the dilaton field as 
\begin{equation}
\Phi = \mu_g^2 r^2,    
\end{equation}
with $r$ being the holographic dimension. The 5D coupling constant $g_5$ can be determined as $g_5^2 = \frac{12\pi^2}{N_c}$ by matching the large-momentum expansion of the vector current $J_\mu^a = \bar{q} \gamma_\mu t^a q$ correlation in AdS/QCD with that from the perturbative calculation \cite{Erlich:2005qh}.
The covariant derivative \(D_M\)  and the field strength tensors $F_{MN}^{L,R}$ are defined as follows:

\begin{equation}\begin{aligned}
D_MX & =\partial_MX-iL_MX+iXR_M, \\
 \\
F_{MN}^L & =\partial_ML_N-\partial_NL_M-i[L_M,L_N], \\
 \\
F_{MN}^R & =\partial_MR_N-\partial_NR_M-i[R_M,R_N],
\end{aligned}\end{equation}
with $L_M,R_M$ being the 5D left- and right-handed gauge potentials, which are dual to the left- and right-handed chiral currents. $V_X(|X|)$ represents the scalar potential, which may generally include a coupling to the dilaton field. Since we consider the degenerate two-flavor case ($m_u=m_d$) only, the expectation of $X$ can be decomposed as $X=\frac{1}{2}\chi I_2$, with $I_2$ the $2\times2$ identity matrix. The factor $1/2$ is introduced to ensure that the kinetic term in $\chi$ is canonical. Then we denote the scalar potential in terms of $\chi$ as $V(\chi)=\text{Tr}[V_X(X)]$. Since we do not consider the case with finite densities or consider the vector perturbations, the relevant part of the action reduces to 
\begin{equation}\label{act-eff}
    S_{5\mathrm{D}} = \int d^5x  \sqrt{g} e^{-\Phi} \big[\frac{1}{2} \chi '^2-V(\chi)\big].
\end{equation}
Here we have assumed that the system is in an equilibrium state, so $\chi$ depends on $r$ only. The overprime $'$ in the above equation denotes differentiation with respect to $r$. Following our previous study \cite{Chelabi:2015cwn}, we choose the scalar potential to be 
\begin{equation}
    V(\chi)=\frac{1}{2} m_5^2(r) \chi^2+\frac{\lambda}{8} \chi^4.
\end{equation}
Here, one can assume a general coupling between the scalar field and the dilaton field, such as $g_2(\phi) \chi^2$ and $g_4(\phi)\chi^4$ with $g_2,g_4$ two coupling functions. In this case, effective $r$-dependent $m_5^2(r)$ and $\lambda(r)$ can be introduced. That is why we treat $m_5^2$ and $\lambda$ as $r$-dependent functions in the above potential. Following  Ref.~\cite{Fang:2016nfj}, we will take 
\begin{equation}
  \text{Model I}: m_5^2(r) = -3 - \mu_c^2 r^2
\end{equation}
and a constant $\lambda$,  which we will call 'Model I'. Here, the leading term $-3$ comes from the AdS/CFT prescription $M^2_5=(d-p)(d+p-4)$ with the dimension $d=3$, and $p=0$ for the scalar $\bar{q}q$ operator. Following Ref.~\cite{Liang:2023lgs}, we will take \begin{equation}
  \text{Model II}: m_5^2(r) =-3[1+\gamma \tanh (\kappa \Phi)]
\end{equation}
and a constant $\lambda$,  which we will call `Model II'. The light meson spectra have been carefully studied in these two models, and the model parameters are fitted as shown in Table \ref{tab:paramters-1} for model I and in Table \ref{tab:paramters-2} for Model II. The quark mass enters the two models through the boundary expansion of $\chi$, and the physical values $m_{phy}$ in Table \ref{tab:paramters-1} and Table \ref{tab:paramters-2} give pion masses around the experimental value (we take $139.6~\rm{MeV}$). Here, the pion masses can be obtained by considering the perturbation of the Goldstone modes (for the details, please refer to Refs.\cite{Erlich:2005qh,Karch:2006pv}). Later, we will study the scaling behavior in these two models. Before that, for a comparison, we also introduce an additional model,
\begin{eqnarray}
\text{Model III:} V(\chi)&=&\frac{1}{2}m_5^2\chi^2+\frac{1}{2}\lambda(r)  \chi^2 \ln(1+b\chi^2),\nonumber\\
    m_5^2(r) &=& -3 - \mu_c^2 r^2,\nonumber\\
    \lambda(r) &=& a_0 + a_1 \tanh(a_2 r^2).
\end{eqnarray}
We will call this model 'Model III'. The model parameters are listed in Table~\ref{tab:paramters-3}. We shall discuss the motivation for this modification after presenting the main numerical results. As can be seen from the setup of the three models, in addition to $\mu_g$ which accounts for linear confinement \cite{Karch:2006pv}, an additional scale is required to describe spontaneous chiral symmetry breaking \cite{Chelabi:2015cwn}.

\begin{table}[htbp]
    \centering
    \caption{	Fitting parameters in Model I. }
    \label{tab:paramters-1}
    \resizebox{0.45\textwidth}{!}{ 
    \begin{tabular}{ccccc}
        \toprule
        Parameters&$m_{phy}(\text{MeV})$ & $ \mu _g (\text{Mev})$ &$ \mu _c (\text{Gev})$&$\lambda$ \\
        \midrule
        Value&3.22 & 0.44&1.45&80 \\
        \bottomrule
    \end{tabular}
    }
\end{table}

\begin{table}[htbp]
    \centering
    \caption{	Fitting parameters in Model II.}
    \label{tab:paramters-2}
    \resizebox{0.45\textwidth}{!}{ 
    \begin{tabular}{cccccc}
        \toprule
        Parameters&$m_{phy}(\text{MeV})$ & $ \mu _g (\text{MeV})$ &$ \gamma$&$\lambda$&$\kappa$ \\
        \midrule
        Value&3.90 & 0.35&6&25&0.85 \\
        \bottomrule
    \end{tabular}
    }
\end{table}

\begin{table}[htbp]
    \centering
    \caption{	Fitting parameters in Model III.}
    \label{tab:paramters-3}
    \resizebox{0.45\textwidth}{!}{ 
    \begin{tabular}{ccccccccc}
        \toprule
        Parameters&$m_{phy}(\text{MeV})$ & $ \mu _g (\text{Mev})$ &$ \mu _c (\text{Gev})$ & $b$ & $a_0$ &$a_1$ &$a_2$ \\
        \midrule
        Value&3.10 & 0.22 & 1.15 & 50 & 20 & 5 & 0.20  \\
        \bottomrule
    \end{tabular}
    }
\end{table}

Besides the scalar fields and their potential, one also needs to take a specific form of the background gravity. In consideration of the symmetries of the four-dimensional theory at finite temperature, the metric ansatz is adopted as follows:
\begin{equation}ds^2 = e^{2A(r)} \left( f(r) dt^2 - dx^2 - \frac{1}{f(r)} dr^2 \right).\end{equation} 
A simplest consideration is to take the AdS-Schwarzschild (AdS-SW) black hole solution with
\begin{equation}
\begin{aligned}
A(r) & =-\ln(r), \\
f(r) & =1-\frac{r^4}{r_h^4}.
\end{aligned}
\end{equation}
Here, \(r_h\) denotes the black hole horizon, at which the blackening factor satisfies \(f(r_h) = 0\). This can be mapped to the temperature through the Hawking temperature defined as
\begin{equation}
T=\left|\frac{f^{\prime}(r_h)}{4\pi}\right|=\frac{1}{\pi r_h}.
\end{equation}

To solve this model, one has to derive the equation of motion for $\chi$ from the action Eq. \ref{act-eff}, which reads
\begin{align}
\chi'' 
+ \left(3A' + \frac{f'}{f} - \phi'\right) \chi'
- \frac{e^{2A}}{f}  V_{,\chi}(\chi)
= 0,
\label{eq:chieom}
\end{align}
with $V_{,\chi}(\chi)\equiv\partial_\chi V(\chi)$.

\begin{figure}[htb]
  \centering

  \begin{subfigure}{0.88\linewidth}
    \includegraphics[width=\linewidth]{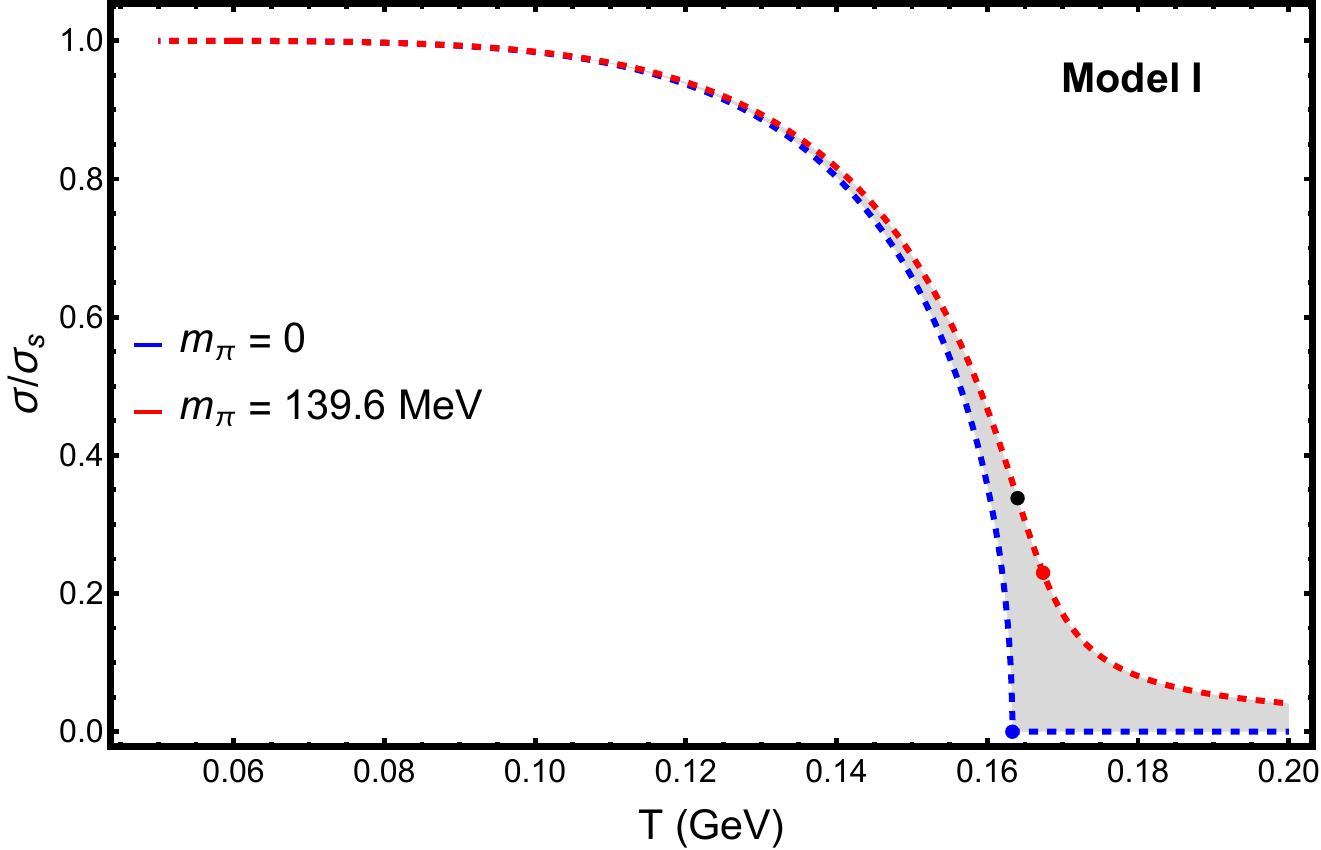}
    \label{fig:sigma-T-1}
  \end{subfigure}

  \begin{subfigure}{0.88\linewidth}
    \includegraphics[width=\linewidth]{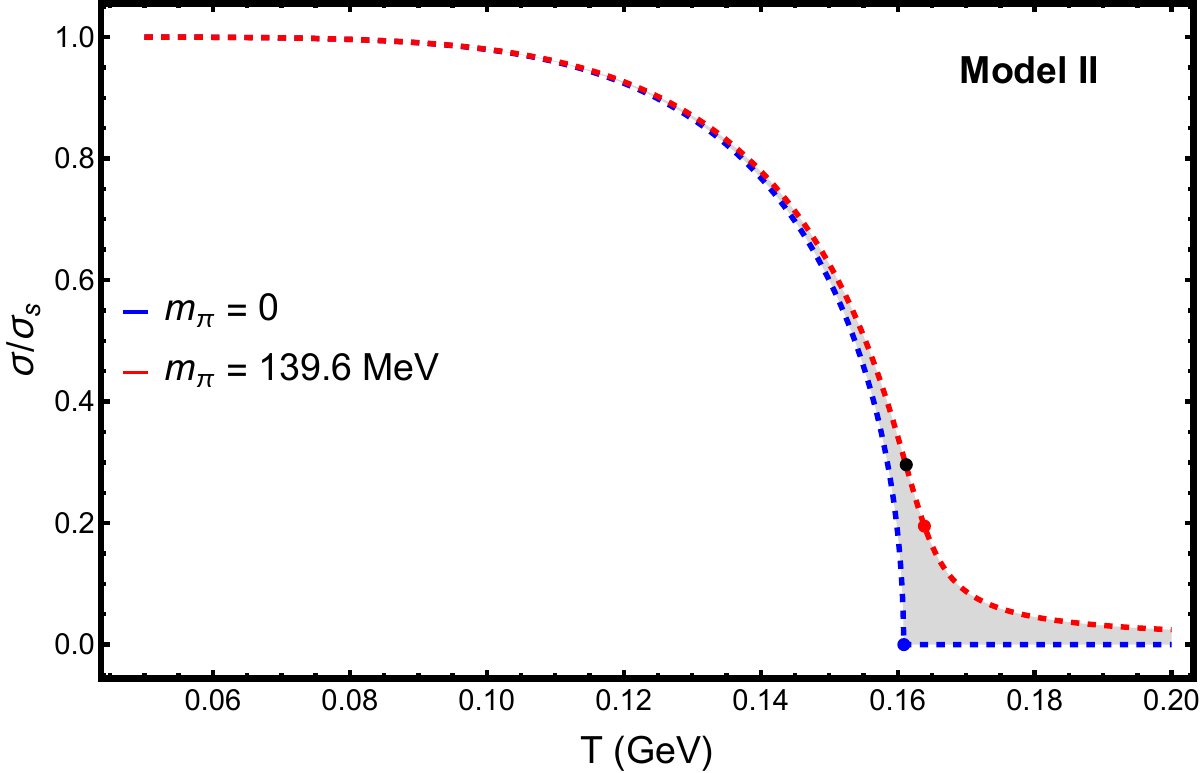}
    \label{fig:sigma-T-2}
  \end{subfigure}
  
  \begin{subfigure}{0.88\linewidth}
    \includegraphics[width=\linewidth]{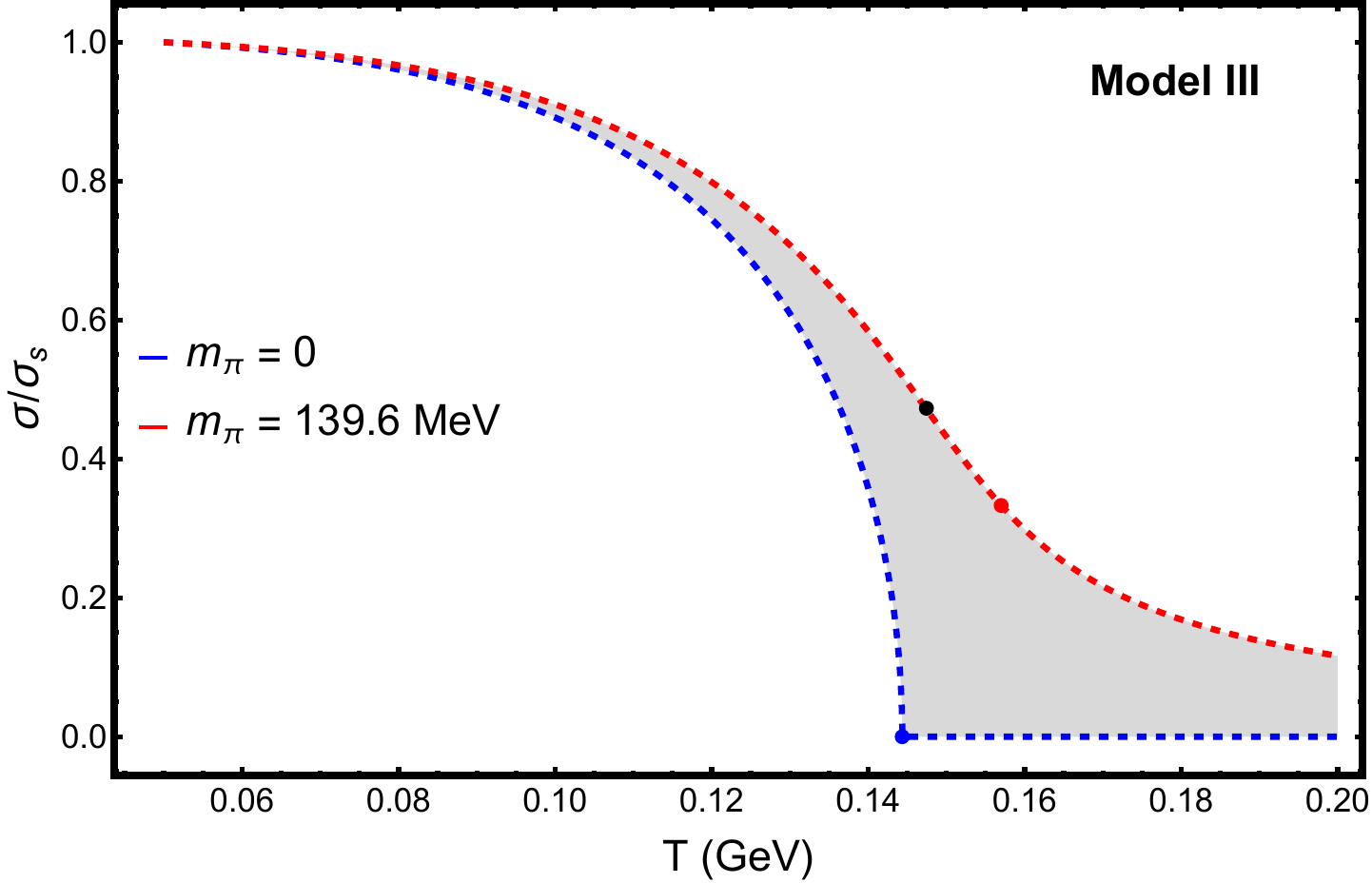}
    \label{fig:sigma-T-3}
  \end{subfigure}
  
  \caption{Chiral phase transitions and the crossover region for different pion masses. The blue line corresponds to the chiral limit \(m_\pi = 0\), while the red line is for the physical pion mass ($m_\pi = 139.6\,\text{MeV}$). The shaded gray area highlights the region between these two curves. Blue dots mark second-order phase-transition points; black and red dots indicate the pseudo-critical temperatures obtained from the maxima of $|d\sigma/dT|$ and the chiral susceptibility, respectively. } \label{sigma-T}
\end{figure}

It is not difficult to derive from the above equation the near boundary ($r=0$) and near horizon ($r=r_h$) expansion of $\chi$ as
\begin{subequations}
\begin{align}
\chi(r\to 0) &= 
m_q r \zeta +...+ \frac{r^3 \sigma}{\zeta} 
+\mathcal{O}(r^4),
 \label{eq:chi0} \\
\chi(r\to r_h) &= 
c_0 - \frac{V^\prime(c_0)}{4r_h}(r - r_h)  +\mathcal{O}(r-r_h)^2,
 \label{eq:chizh}
\end{align}
\end{subequations}
where $m_q$ and $\sigma \equiv \langle \bar{q} q \rangle$  are two independent integration constants in the UV region, dual to the quark mass and the chiral condensate, respectively. $c_0$ is an integration constant in the IR region and $c_0\equiv\chi(r_h)$.
In Ref.~\cite{Cherman:2008eh}, $\zeta$ is identified as a normalization constant with the value $\zeta = \frac{\sqrt{N_c}}{2\pi}$, which is obtained by mapping the correlation of the scalar operator to that from the 4D perturbative calculation. Here, we will take $N_c = 3$.
One of the integration constants in the IR region is discarded due to its association with a divergent solution at $r_h$. Thus, $c_0$ is kept as the only coefficient corresponding to a regular solution at the horizon.

This study focuses on the scaling behavior of the chiral phase transition temperature. By utilizing the asymptotic expansions of the solutions near the UV boundary $r = 0$ and the IR horizon  $r = r_h$  as provided in Eqs.\eqref{eq:chi0} and \eqref{eq:chizh}, the boundary-valued problem can be reformulated as an iterative problem following the procedure established in our previous work~\cite{Chelabi:2015cwn,Cao:2021tcr}. This enables the application of the “shooting method”~\cite{boyd2001chebyshev} to extract the chiral condensate $\sigma$.

Inserting the parameters in Tables \ref{tab:paramters-1}, \ref{tab:paramters-2} and \ref{tab:paramters-3} into the three models, one can obtain the chiral condensate as a function of temperature. Furthermore, since the quark mass appears in the boundary condition, it is also possible to vary its value, and correspondingly, one gets the function $\sigma(m_q, T)$. We plot the results from the three models in Fig. \ref{sigma-T}. In the upper, middle, and lower panels of the figure, the results of Model I, II and III are presented, respectively. For each model, we show two different cases, with a physical quark mass and in the chiral limit. For the physical case, we tune the quark mass to get a pion mass of around $139.6\rm{MeV}$, while in the chiral limit, we take $m_q=0$ and correspondingly $m_\pi=0$ for its Goldstone nature. As can be seen from the figures, in the chiral limit, all three models predict second order phase transitions. The vacuum values of the condensate $\sigma_s$ and the second order transition temperature are $\sigma_s= 0.01475 ~\rm{GeV}^3, T_{c0}= 163.2~\rm{MeV}$ for model I, $\sigma_s= 0.02749~\rm{GeV}^3, T_{c0}=160.9 ~\rm{MeV}$ for model II, and $\sigma_s=0.002175 ~\rm{GeV}^3, T_{c0}=144.4 ~\rm{MeV}$ for model III. In our calculation, it is found that any finite quark masses will drive the phase transition from second order to continuous crossover for all three models. It can be seen from the figures that at the physical quark masses the crossovers in Models I and II appear as a very sharp and fast transition. 

Compared to these two models, Model III yields a significantly broader crossover. This behavior can be attributed to the modification of the scalar potential. As shown in Ref. \cite{Chen:2018msc}, in this model a larger quartic coupling suppresses the chiral condensate at low temperatures. We therefore introduce a quartic coefficient that depends on the coordinate $r$ (its $r$-dependence arises from coupling to the dilaton), and make it grow with increasing $r$. Consequently, the low-temperature condensate is suppressed more strongly than at high temperatures. Additionally, the logarithmic term softens the potential at large $\chi$, causing the condensate to approach its $T = 0$ limit more gradually. The interplay of these two effects broadens the temperature profile of the condensate. For the physical quark masses, the pseudo-critical temperatures extracted from the maxima of $|d\sigma/dT|$ are $T_c=164.0 \rm{MeV}$ for Model I, $T_c=161.2 \rm{MeV}$ for Model II, and $T_c=147.4 \rm{MeV}$ for Model III (black dots), while those from the maxima of the chiral susceptibility are $T_c=167.4 \rm{MeV}$ for Model I,  $T_c=163.9\rm{MeV}$ for Model II, and $T_c=157.0\rm{MeV}$ for Model III (red dots).

\section{ Scaling functions in the holographic model}\label{secion:scaling funciton}

As discussed in the last section, with arbitrarily small quark mass, the chiral phase transition turns into a crossover. When the transition point lies sufficiently close to the critical point, it can be strongly influenced by the critical point. A scaling window and universal scaling behavior appear in this region, which might be relevant to the search for the critical end point (CEP) of the QCD phase diagram if the CEP belongs to the same universality class. In this context, it has been verified that the chiral condensate near the critical point obeys the following scaling laws:
\begin{equation}
\sigma(T,m_q=0)\sim(T_{c0}-T)^\beta, \sigma(T_{c0},m_q)\sim m_q^{1/\delta},
\end{equation}
with $\beta=1/2$ and $\delta=3$ in the two-flavor limit \cite{Chen:2018msc}. However, these scaling laws capture the critical behavior  only along certain directions---namely, holding $m_q$ fixed for $\beta$ and $T$ fixed for $\delta$. To get more information about how the critical point governs the crossover region, one has to extract the scaling functions. This section is therefore devoted to their determination.

\subsection{Scaling functions}\label{scalin funtion}

The theoretical derivation of the universal scaling near a critical point can be easily found in a standard textbook~\cite{KADANOFF:1967qhc} of statistical physics. Taking a spin system as an example, the order parameter or the magnetization $M$ obeys the scaling law
\begin{align}
M=h^{1/\delta}f_{G}(z),
\label{eq:M-fG}
\end{align}
where \( f_G(z) \) is the scaling function, $h$ is the source of the magnetization, and $z$ is a scaling parameter defined as
\begin{align}
z={t} / h^{1 / \Delta},  t=(T_{c0}-T)/T_{c0},
\label{eq:z}
\end{align}
with $ \Delta = \beta\delta$.
If one takes $T=T_{c0}$ or $h=0$, the scaling law will be reduced to 
\begin{subequations}
\begin{align}
M\left(t = 0\right)=h^{1/\delta }, \\
M(h=0)= t^{\beta } ,
\label{eq:MT}
\end{align}
\label{eq:solm0}
\end{subequations}
which implies the conditions for $f_G$
\begin{equation}
f_G(0)=1, \quad \text { and } \quad f_G(z) \underset{z \rightarrow-\infty}{\rightarrow}(-z)^\beta .
\label{eq:fg0}
\end{equation}

\begin{figure*}[htb]
  \centering
  \includegraphics[width=0.95\textwidth]{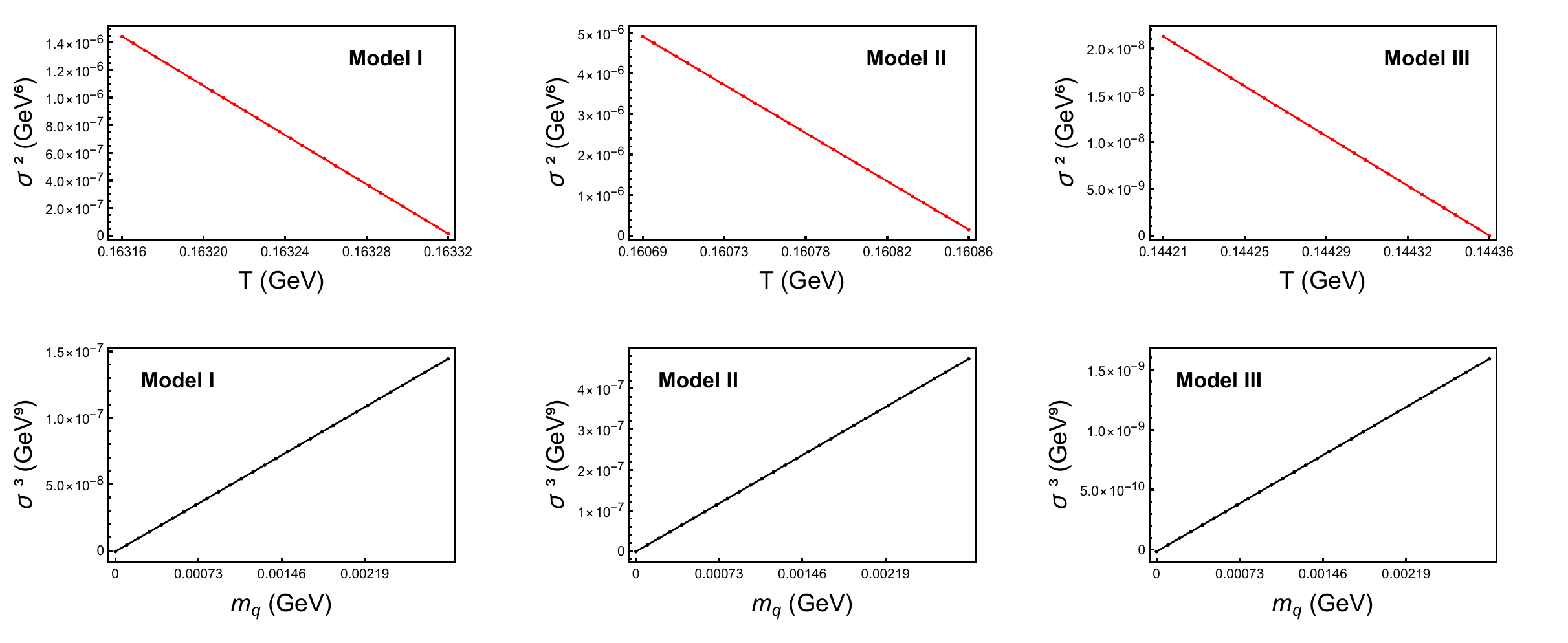}
  \caption{Critical scaling behavior near the critical point across different models. Red and black symbols are numerical data from the three models; lines are fits.}
  \label{fig:βδ-AB}
\end{figure*}

In the current model, the critical point locates at $T=T_{c0},~ m_q=0$. It can be numerically checked that the order parameter $\sigma$ in all three models obeys the following scaling law
\begin{subequations}
\begin{align}
\sigma \left(T =T_c,m_q\right)=A m_q^{1/\delta },
\label{eq:sgT} \\
\sigma (T,m_q = 0)=B\left(T_c-T\right){}^{\beta }.
\label{eq:Mm}
\end{align}
\end{subequations}
with \( A \) and \( B \) two coefficients, which can be numerically extracted.
One can take $m_q=0$ and vary $T$ around $T_{c0}$ to get the near critical temperature scaling. The results for the three models are given in Fig. \ref{fig:βδ-AB}. It is shown that $\sigma^2$ is almost a straight line as a function of $T$, indicating $\beta=1/2$. By fitting the slope of the straight lines one can get the coefficient $A$ as listed in Table~\ref{tab:AB}. Similarly, one can take $T=T_{c0}$ and vary $m_q$ around $m_q=0$. It is found that in this case $\sigma^3$ is almost lying in a straight line as a function of $m_q$. Thus, one gets $\delta=3$. The coefficient $B$ can be obtained numerically, and we list the result in Table~\ref{tab:AB}. In fact, those quantities can also be extracted analytically.  Please refer to \cite{Chen:2018msc} for the details.

\begin{table}[htbp]
  \centering
  \begin{tabular}{|c|c|c|}
    \hline
\textbf{Model} & \(A\ (\mathrm{GeV}^{8/3})\) & \(B\ (\mathrm{GeV}^{2})\) \\
\hline
Model I   & 0.0367383 & 0.0935534 \\
Model II  & 0.0545202 & 0.172141 \\
Model III & 0.00818794 & 0.0121398 \\
    \hline
  \end{tabular}
  \caption{The extracted values of \( A \) and \( B \) as determined by various models.
}
  \label{tab:AB}
\end{table}

To obtain the more general scaling function $f_{G}(z)$, we follow the convention of \cite{engels2012scaling} and define the magnetization as $M={\sigma }/{\sigma _0}$, the source $h$ as  $h = {m_q}/{m_0}$ , and $ t = {(T - T_{c0})}/{T_{c0}}$ . After that, we get 
\begin{subequations}
\begin{align}
M\left(t=0,h\right)=\frac{A h^{1/\delta } m_0^{1/\delta }}{\sigma _0},
\label{eq:sgT} \\
M(t,h=0)=\frac{B t^{\beta }T_{c0}^{\beta }}{\sigma _0}.
\label{eq:MT}
\end{align}
\label{eq:M-eq}
\end{subequations}
Then we get the scaling function \( f_G(z) \) as $f_{G}(z)=M/h^{1/\delta}$. By substituting the condition from Eq.~\ref{eq:solm0} back into Eq.~\ref{eq:M-eq}, we can solve for $ m_0 $ and $ \sigma_0 $, and then get the numerical results for $f_G(z)$. We will discuss the numerical results in the next subsection.

In the literature, another scaling function, \( f_\chi(z) \),  is widely studied. It is defined from the chiral susceptibility $\chi_{M}$ as
\begin{align}
     \chi _ {M}  =  \frac {\partial M}{\partial h}  =  \frac {h^ {1/\delta-1 }}{m_ {0}}  f_ {\chi}  (z) ,
     \label{eq:Xm-fchi}
\end{align} 
which can be extracted from the near-critical values of chiral susceptibility in terms of the chiral condensate, 
\begin{equation}
\chi_{M}=\frac{m_0}{\sigma_0}\chi_\sigma=\frac{m_0}{\sigma_0}\frac{\partial \sigma}{\partial m_q}.
\end{equation}
Here, although $\chi_\sigma$ can be extracted by numerically differentiating $\sigma$ with respect to $m_q$, it can be directly extracted by performing a perturbative analysis of Eq.~\ref{eq:chieom}. If one gets a solution $\chi_0$ at a given temperature $T$ and $m$, a small variation $\delta m_q$ in the quark mass will induce a variation $\delta\chi$ in $\chi_0$. Then, it is easy to derive the equation of motion for $\delta \chi$ as
\begin{equation}
\delta \chi'' + \left(3A' + \frac{f'}{f} - \phi'\right) {\delta \chi}'- \frac{e^{2A}}{f}V_{,\chi\chi}(\chi_0) \delta\chi = 0.
\end{equation}
As in the above discussion, we can obtain the expansion of $\delta\chi$ at $r = 0$ and $r = r_h$ as 
\begin{subequations}
\begin{equation}
\begin{aligned}
\delta\chi(r \to 0) =\; &
 \delta m_q  \zeta +...+ \frac{\delta \sigma}{\zeta} r^3+\mathcal{O}(r)^2 ,
\end{aligned}
\label{eq:dchi0}
\end{equation}
\begin{equation}
\begin{aligned}
\delta\chi(r \to r_h) =\; &
1+  \frac{1}{4r_h}(r - r_h) +\mathcal{O}(r-r_h)^2.
\end{aligned}
\label{eq:dchizh}
\end{equation}
\end{subequations}
Then, using the same `shooting method' as in Sec.\ref{section:model}, one can obtain $\delta\sigma$ as a function of $\delta m_q$ for given temperatures. Thus, the chiral susceptibility can be obtained as $\chi_\sigma=\delta \sigma/\delta m_q$. We will discuss the numerical results later. Before that, we note that the scaling functions $f_G$ and $f_\chi$ have a general relation. Performing the differentiation with respect to $m_q$, it is not difficult to get
\begin{align}
    f_\chi(z)=\frac1\delta {f_G(z)} -\frac z{\beta\delta} f_G^{\prime}(z).
    \label{eq:fchi-fG}
\end{align}

Considering the case \( T = T_{c0} \), Eqs.~\ref{eq:M-fG} and ~\ref{eq:Xm-fchi}  yield that the ratio between the two scaling functions approaches a constant near the critical point, i.e.
\begin{align}
    \frac{f_{\chi}(0)}{f_{G}(0)}=\frac{1}{\delta}.
    \label{eq:fchi-0}
\end{align}
The above relations are quite general for different systems. Thus, a self-consistent holographic model should produce results obeying those constraints. We will check the numerical results from the soft-wall model in the next section.

\begin{figure}[htb]
  \centering

  \begin{subfigure}{\linewidth}
    \includegraphics[height=0.18\textheight]{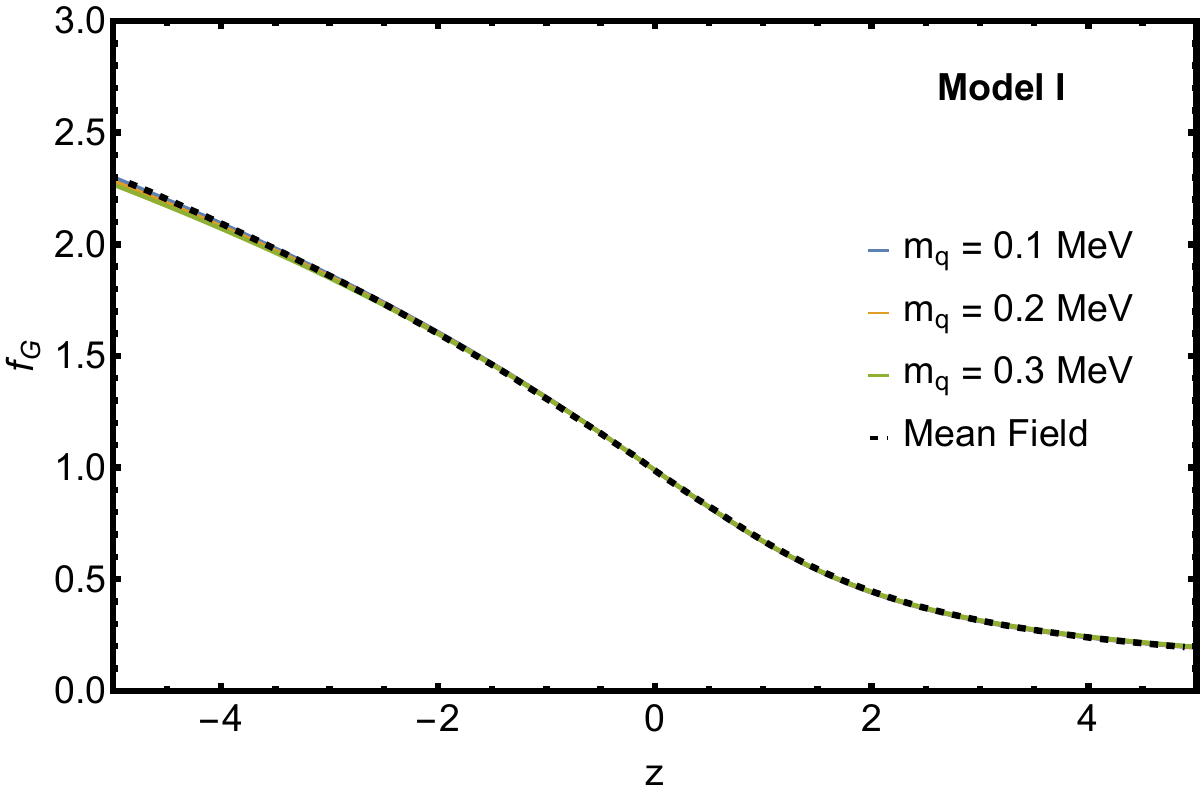}

    \label{fig:fg-2}
  \end{subfigure}
  \vskip 1em

  \begin{subfigure}{\linewidth}
    \includegraphics[height=0.18\textheight]{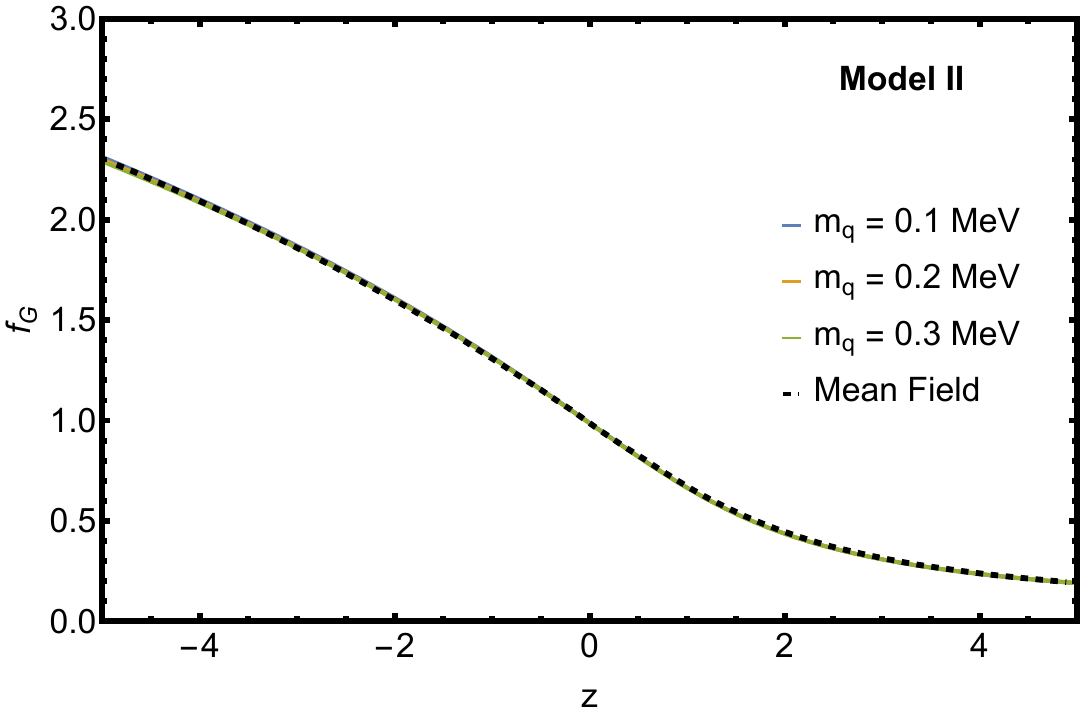}
    \label{fig:fg-3}
  \end{subfigure}
  \vskip 1em
  
  \begin{subfigure}{\linewidth}
    \includegraphics[height=0.18\textheight]{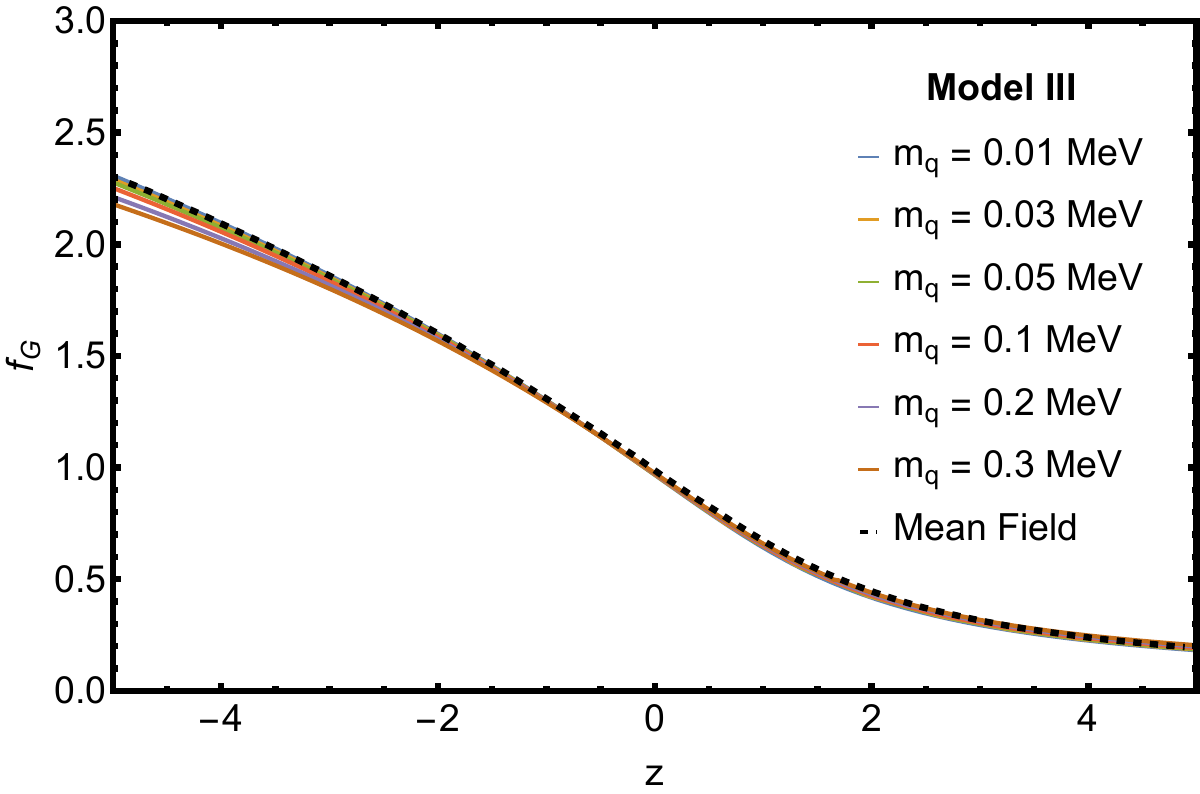}
    \label{fig:fg-1}
  \end{subfigure}  
  \vskip 1em
  
  \caption{Scaling function $f_G(z)$ for the three holographic models at several finite quark masses, together with the mean-field result (black solid line) taken from Ref.~\cite{Grossi:2021gqi-meanfied}.}
  \label{fig:fG-Model}
\end{figure}

\begin{figure}[htb]
  \centering
  \includegraphics[width=0.48\textwidth]{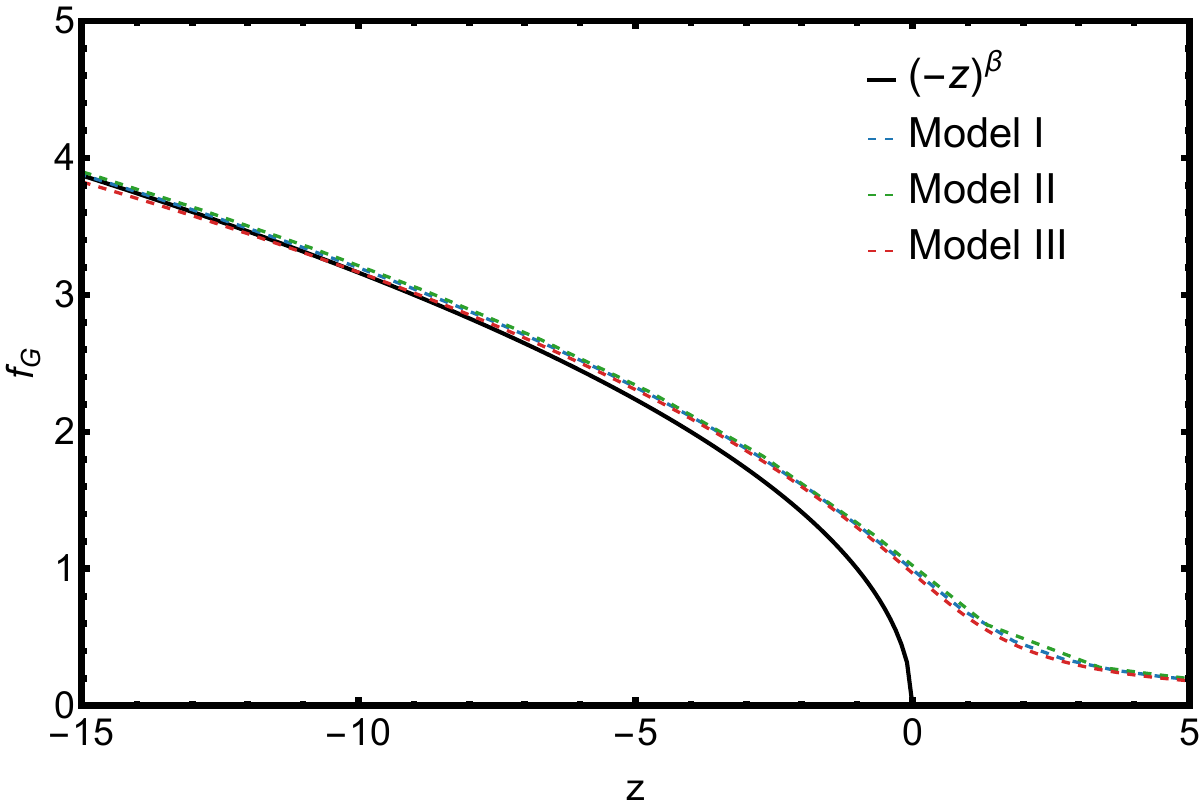}
  \caption{The asymptotic fitting of the scaling function  $f_G$  to  $-z^\beta$  as  $z \to -\infty$. The quark mass is set to  $m_q = 0.01\,\text{MeV}$ for all three models.
}
  \label{fig:-t-data}
\end{figure}

\subsection{Numerical scaling functions from the soft-wall AdS/QCD models}\label{scalin funtion}

With the preparations described above, we numerically extract the scaling functions in the soft-wall AdS/QCD models.  First, we obtain the numerical data for the chiral condensate $\sigma$ near the critical point ($T$ near $T_{c0}$ and $m$ around $0$). Then from Eqs.~\eqref{eq:M-fG} and \eqref{eq:sgT}, we can extract the results of $f_G(z)$. The numerical results of the three models and the comparison with the mean-field result from Ref.~\cite{Grossi:2021gqi-meanfied} are presented in Fig. \ref{fig:fG-Model}. For each case, we take $m_q=0.1,~0.2,~0.3 ~\rm{MeV}$ as examples. It can be seen that for all models the scaling function $f_G(z)$ approaches the mean-field result (the black solid lines) from \cite{Grossi:2021gqi-meanfied} when the quark mass approaches the chiral limit. It can be checked that $f_G(0)=1$ for all models, consistent with the definition. In the range of $-2<z<5$, all three models merge with mean-field results together. But in the range of $-5<z<-2$, the results in Model I and II converge much faster to the mean-field, while for Model III it is slower and agrees with the mean-field result only when $m_q$ is of order $10^{-2}~\rm{MeV}$ $f_G(z)$ (that is why we add $m_q=0.01,0.03,0.05 ~\rm{MeV}$ for Model III). It is not difficult to understand this behavior, as the increase in $m_q$ will drive the system away from the critical point, resulting in a violation of the scaling behavior. However, it appears that the scaling window for Model III is smaller than the other two models. Later, we will see the impact of this in the scaling of the pseudo-critical temperatures.

\begin{figure}[htb]
  \centering
  \includegraphics[width=0.48\textwidth]{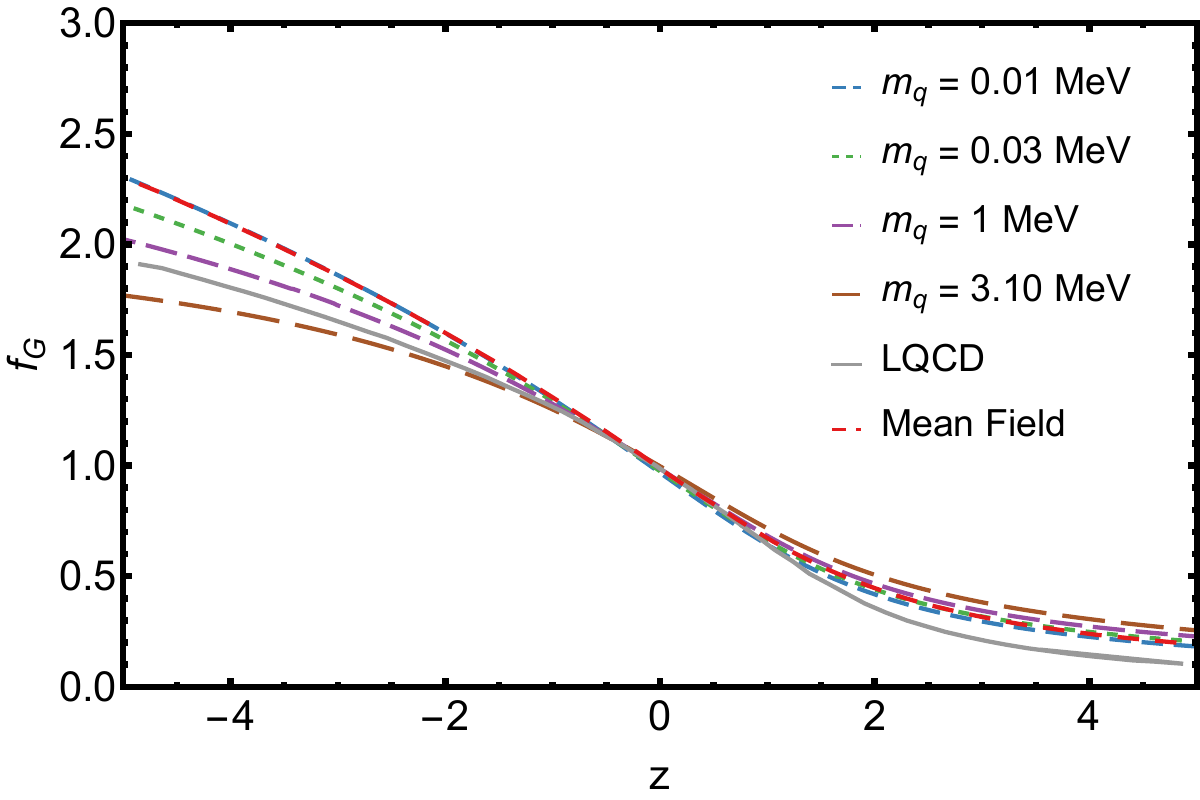}
  \caption{A comparison of the scaling function obtained from Model III with those derived from the mean-field approximation \cite{Grossi:2021gqi-meanfied} (the red dashed line) and lattice QCD simulations for $O(4)$ model \cite{Engels:2011km} (the black solid line).}
  \label{fig:fg-Lqcd-mf}
\end{figure}

As a check of the consistency of the model, we also examine the behavior of $f_G(z)$ as $z$ approaches $-\infty$. It is shown in Fig. \ref{fig:-t-data} that when $z$ decreases to $-8$, in the three models, $f_G(z)$ can be well described by $(-z)^{1/2}$. From these calculations, we confirm that the full critical scaling of the soft-wall AdS/QCD models is governed by the mean-field scaling. To construct a more realistic model, we also compare the scaling function in Model III with the lattice results for the $O(4)$ model in Fig.\ref{fig:fg-Lqcd-mf}. From the figure, obvious deviations of the model results from the lattice results appear in the range of $z<-1$ and $z>1$. Though the scaling function is defined at the critical point, for later discussion, we also present a direct extension through Eqs.~\eqref{eq:M-fG} and~\eqref{eq:sgT} to $m_q=0.01,~0.03,~1 \rm{MeV}$ and the physical point. Interestingly, the extended scaling function bends towards the lattice results in the region of negative $z$. As the quark mass slightly increases, part of the scaling function gradually shifts toward the  $O(4)$  behavior observed in lattice QCD~\cite{Engels:2011km, Engels:2014bra}. By comparison, the other two soft-wall models yield scaling functions that remain valid over a broader range as the quark mass changes. This might help improve the temperature scaling within the framework of the soft-wall model, which can be seen in the later discussion.

\begin{figure}[htb]
  \centering
  \includegraphics[width=0.48\textwidth]{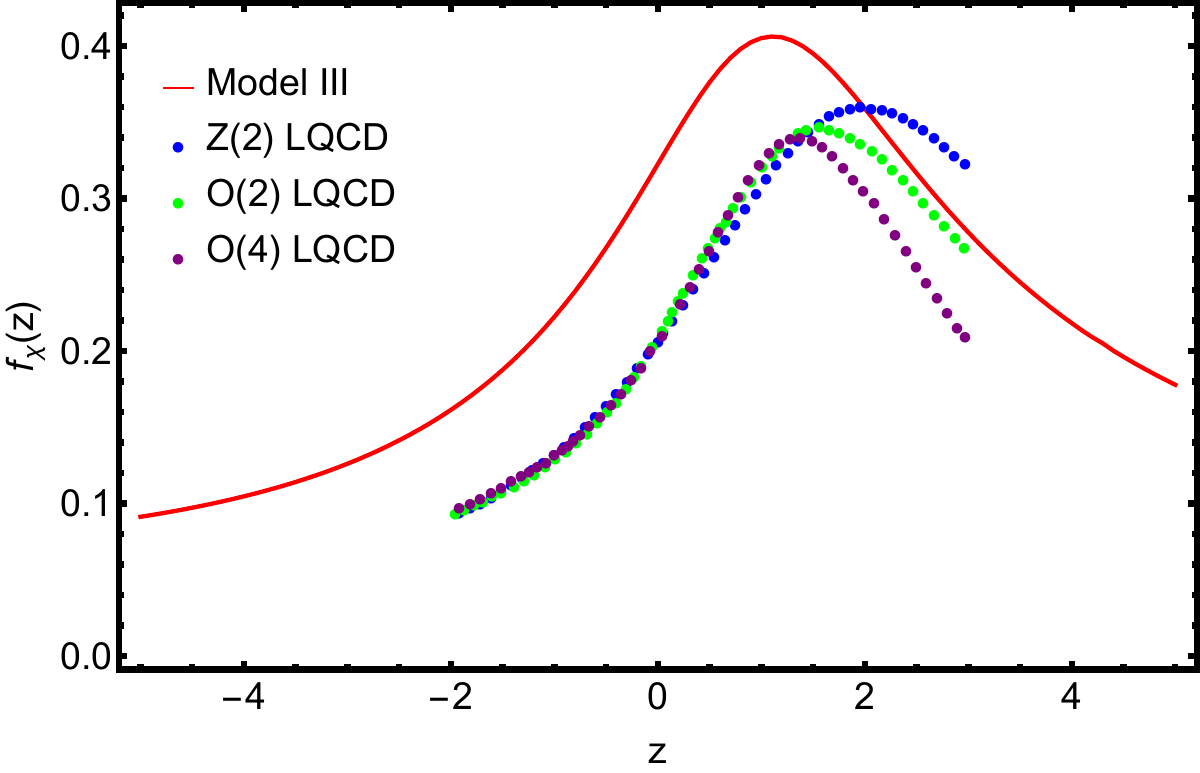}
  \caption{A comparison of the scaling function  $f_\chi(z)$ from Model III, with the corresponding results derived from lattice QCD in the $ O(2)$, $Z(2)$ , and $ O(4) $ universality classes \cite{HotQCD:2019xnw}.}
  \label{fig:fg-Lqcd-mf}
\end{figure}

For the scaling function $f_\chi$, as mentioned in the previous section, it can be obtained in two different ways. Firstly, we can directly extract it through the general relation Eq.\eqref{eq:fchi-fG}.  The results in Model III are given in Fig. \ref{fig:fg-Lqcd-mf}, compared with the lattice simulation of Ref.~\cite{HotQCD:2019xnw}. From the figure, we obtain $f_\chi\simeq0.322$. From Eq. \ref{eq:fchi-0}, this value should be $f_\chi(0)=\frac{1}{\delta} f_G(0)=\frac{1}{3}$. The deviation from this exact value is mainly from the numerical errors, due to the complexity of directly taking $m=0$ (instead, we take a tiny but finite value). It can also be seen that a peak with its center at $z=1.115,f^{\text{max}}_\chi=0.406$ appears in the curve for $f_\chi$. The ratio $f_\chi(z=0)/f_{\chi,\text{max}} $ is approximately equal to $79\%$. Later, we will see an interesting connection between this ratio and the temperature scaling behavior. Since the holographic computation is restricted to the mean-field approximation, its prediction for $f_\chi$  cannot quantitatively agree with any of the lattice results obtained for O(4), O(2), or Z(2) universality classes. The underlying scaling relations are nevertheless respected; all differences stem solely from the distinct critical exponents.

\begin{figure}[htb]
  \centering

  \begin{subfigure}{\linewidth}
    \includegraphics[height=0.18\textheight]{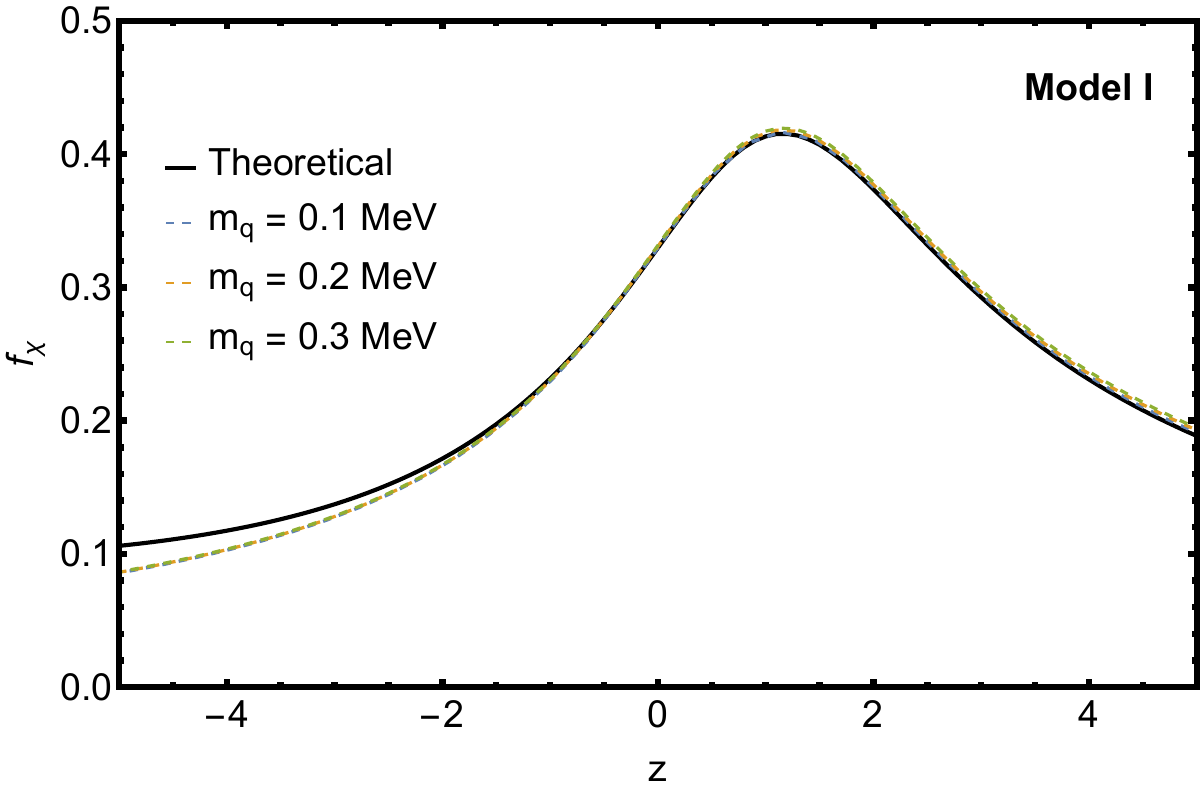}

    \label{fig:fchi-1}
  \end{subfigure}
  \vskip 1em

  \begin{subfigure}{\linewidth}
    \includegraphics[height=0.18\textheight]{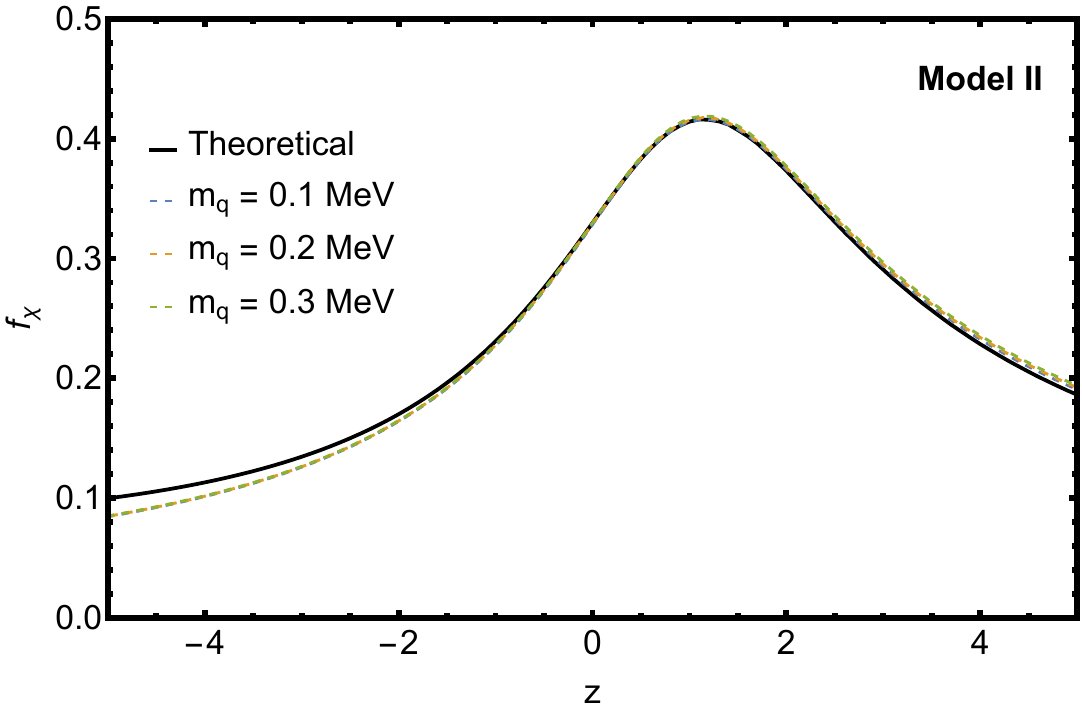}

    \label{fig:fchi-2}
  \end{subfigure}
  \vskip 1em

  \begin{subfigure}{\linewidth}
    \includegraphics[height=0.18\textheight]{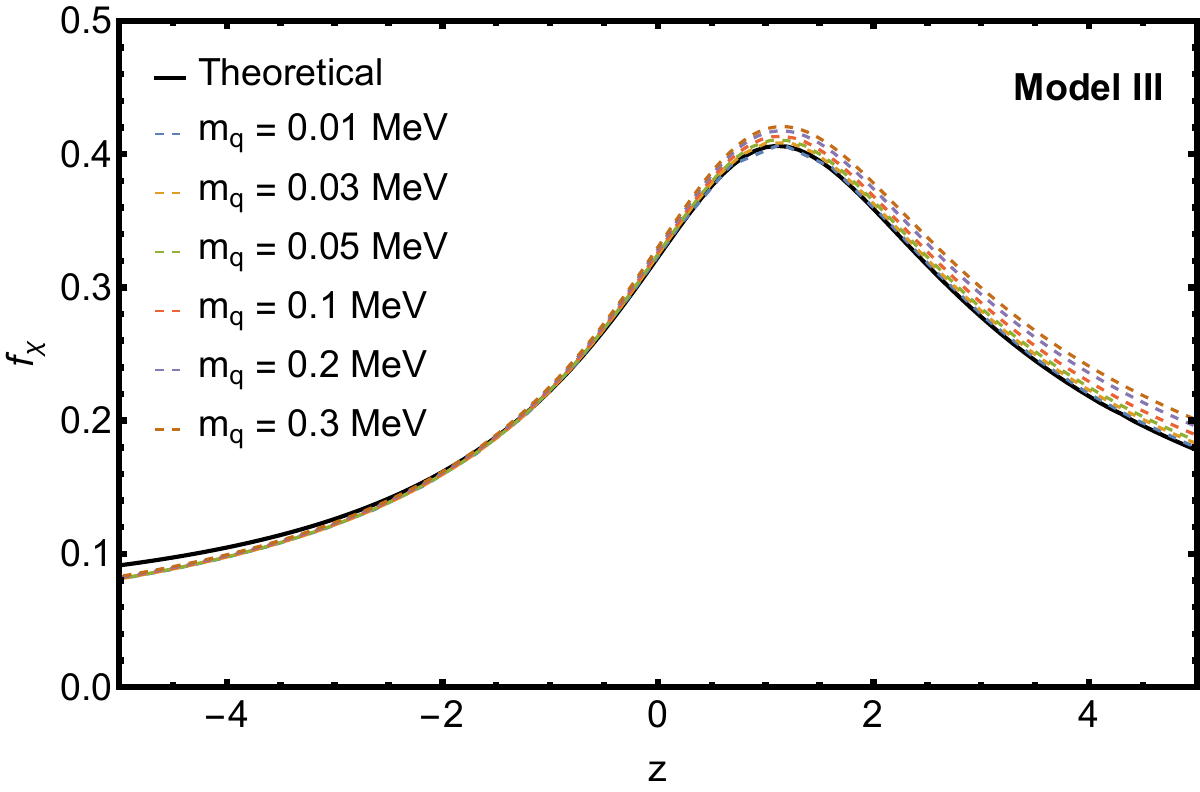}

    \label{fig:fchi-3}
  \end{subfigure}

  \caption{The dependence of the scaling function $f_\chi$ on the quark mass. The solid curve denotes the theoretical prediction extracted from $f_G$ through Eq.\eqref{eq:fchi-fG}, whereas the dashed curves illustrate the scaling functions for various quark masses.
}
  \label{fig:fchi-Model}
\end{figure}

We can also obtain $f_\chi$ from the chiral susceptibility through Eq.\eqref{eq:Xm-fchi}. Fig.~\ref{fig:fchi-Model} presents the computed $f_\chi$  from various models in this way, compared to the theoretical predictions obtained from Eq.~\eqref{eq:fchi-fG}.  From the figure, we can see that the results obtained from the two methods are almost the same when the quark masses are very small. Thus, the general relation between $f_G$ and $f_\chi$ is satisfied in the soft-wall model. Notably, Model III displays a stronger dependence on the quark mass, while Model I and Model II exhibit relatively minor changes in their scaling functions. This suggests that, under the mean-field approximation, variations among models may lead to different scaling regions. 
 
In a short summary, all the soft-wall models obey the same scaling behavior. The obtained scaling functions are the same for various models. However, when extended to finite quark masses, the scaling functions might show a slight dependence on the quark mass, and the scaling region might be different for different models.

\section{Tc scaling behavior under the chiral phase transition}\label{section:tc scaling}

As already mentioned in the above sections, by taking a tiny but finite quark mass, the chiral phase transition turns from a second-order transition into a continuous crossover. Although the continuous crossover is not a real phase transition, it is a smooth and rapid change between two different phases, and one can define a pseudo-critical temperature in different ways. Generally, the pseudo-critical temperature also obeys a certain kind of scaling law. In this section, we will investigate the scaling behavior of the transition temperature.

\subsection{Chiral order parameters}

Several kinds of chiral order parameters can be used in a pseudo-transition. For the holographic calculation, two of them are suitable for extracting the transition temperature. One of them is defined from the peak of the derivative with temperature $|d\sigma/dT|$, while the other is from the peak of the chiral susceptibility defined as $\chi_\sigma=  \frac{\partial \sigma}{\partial m_q}$.

\begin{figure}[htb]
  \centering
  \includegraphics[width=0.48\textwidth]{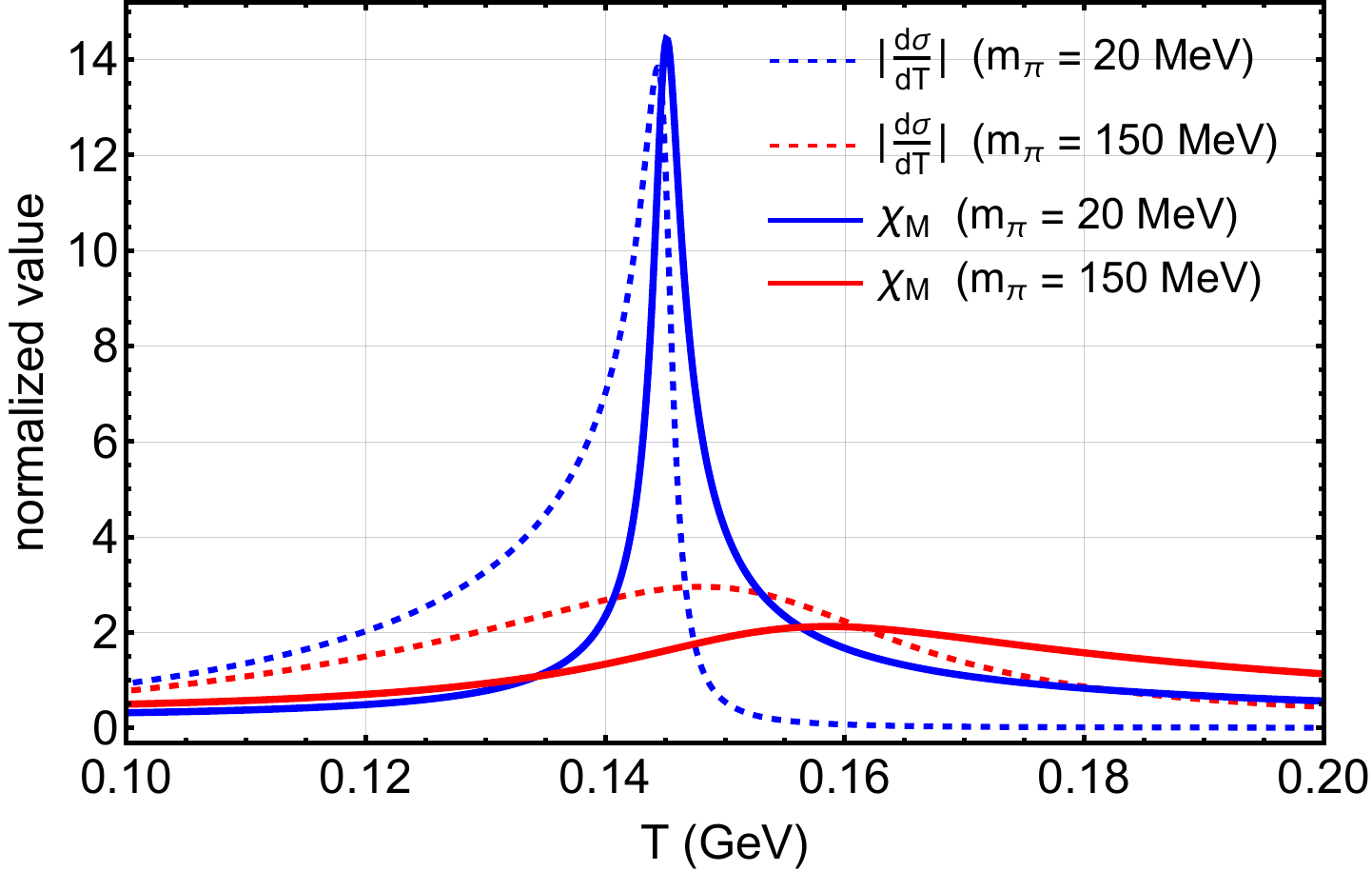}
  \caption{The critical temperatures $T_c$ from chiral susceptibility and chiral condensate, with $m_\pi = 150\,\text{MeV}$ and $m_\pi = 20\,\text{MeV}$. The normalized value denotes each susceptibility scaled by the mean of all susceptibilities.
}
  \label{fig:Xm-dσdT}
\end{figure}

According to Ref.~\cite{Gao:2016qkh}, although the chiral order parameters differ, the extracted pseudo-critical temperatures differ only slightly, whereas Ref.~\cite{Aarts:2020vyb} shows that different chiral order parameters can lead to much larger discrepancies in pseudo-critical temperature. To study the impact of the chiral order parameters, we take two different cases in Model III with $m_q=0.059 \rm{MeV}, m_\pi=20 \rm{MeV}$ and $m_q=3.63\rm{MeV}, m_\pi=150\rm{MeV}$ as examples. As shown in Fig. \ref{fig:Xm-dσdT}, we can extract $T_c$ by identifying the temperatures corresponding to the peaks of the two different chiral order parameters. With $m_q=0.059 \rm{MeV}$, as shown in the blue curves, the peak of $d\sigma/dT$ locates at $T_{c,\sigma}=0.1444 \rm{GeV}$ with a height of $H_\sigma=13.83.$, which are very close to $T_{c,\chi}=0.1451 \rm{GeV}, H_{\chi}=14.42$. As for $m_q=3.63\rm{MeV}$, both the peak location and its height obtain larger differences, with $T_{c,\sigma}=0.1481\rm{GeV}, H_\sigma=2.96$ and $ T_{c,\chi}=0.1586 \rm{GeV}, H_\chi=2.13$. 

\begin{figure}[htb]
  \centering
  \includegraphics[width=0.48\textwidth]{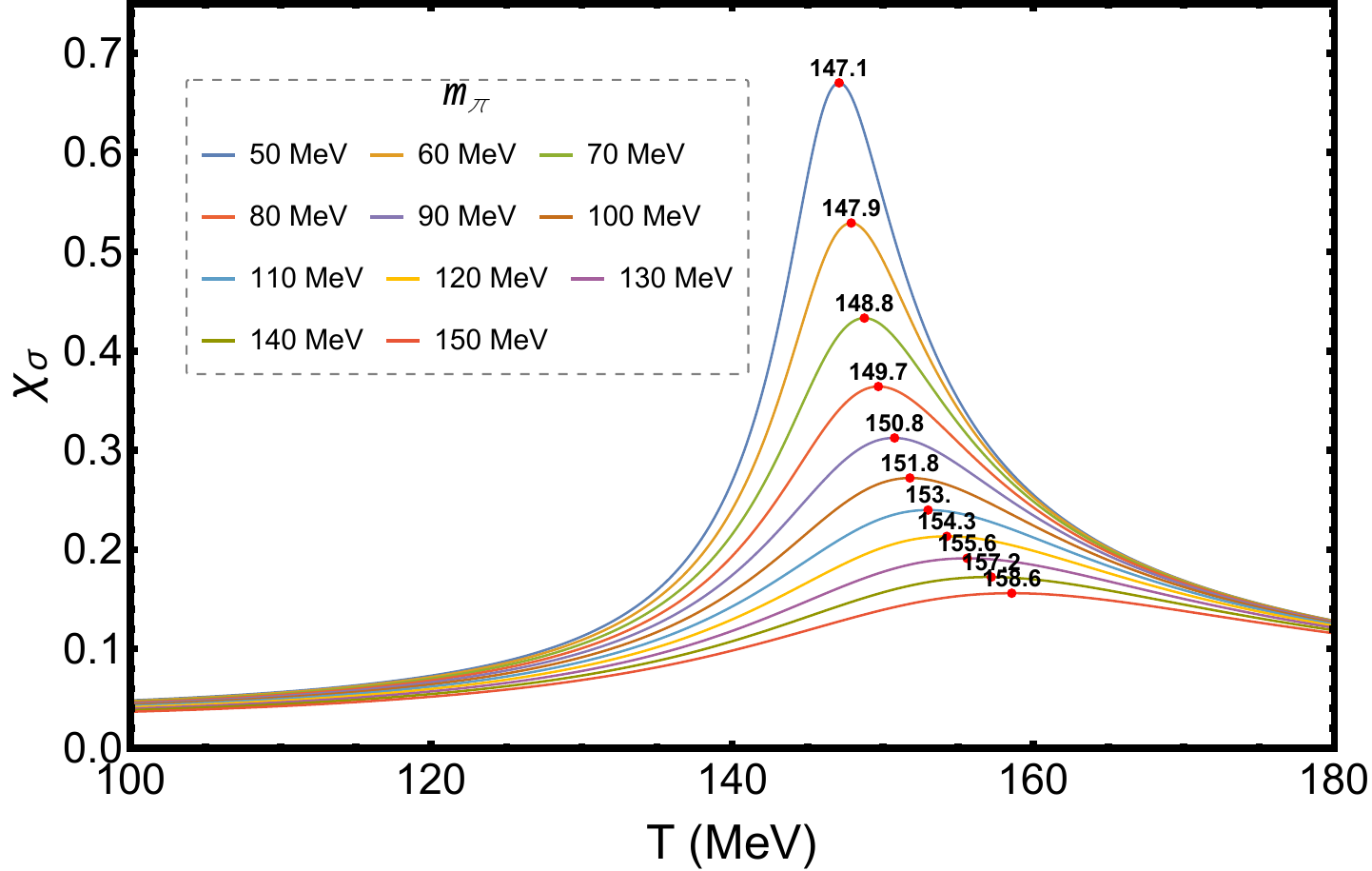}
  \caption{Chiral susceptibility for various pion masses in Model III.}
  \label{fig:XM-0-150}
\end{figure}

\begin{figure}[htb]
  \centering
  \includegraphics[width=0.48\textwidth]{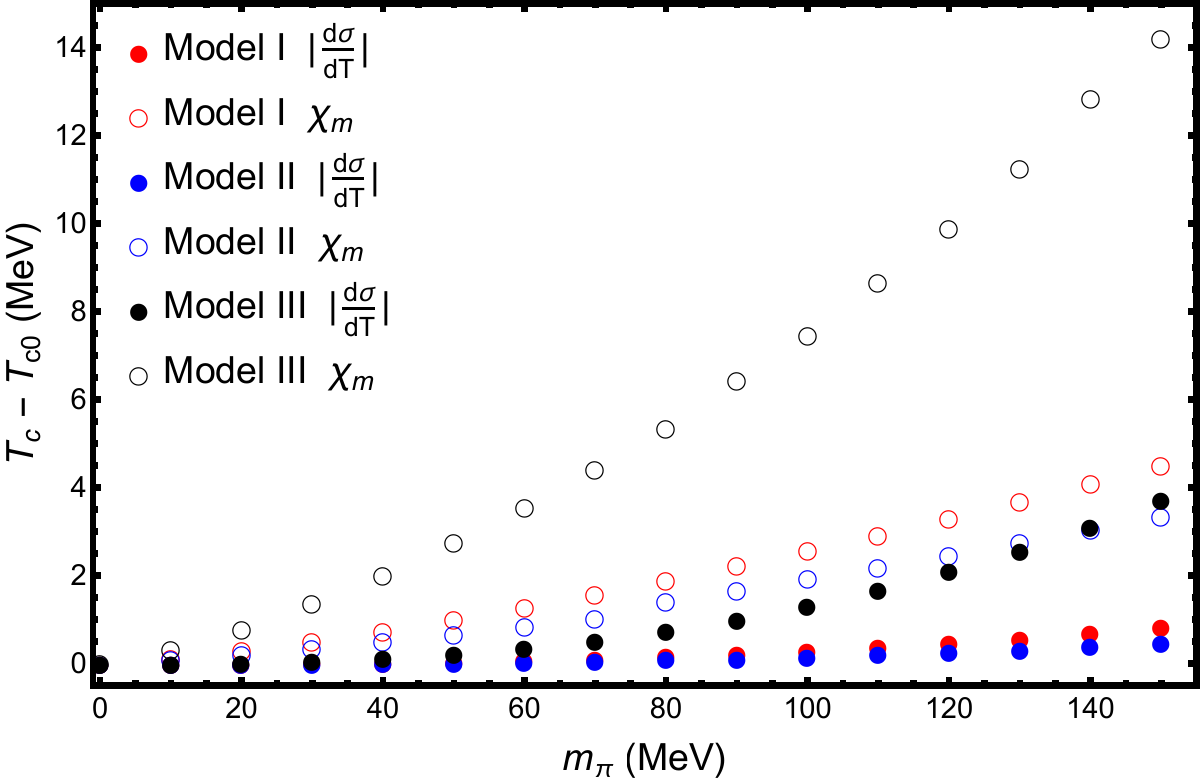}
  \caption{The pseudo-pion mass varies with different chiral order parameters and the temperature of the phase transition. The open circles represent the critical temperatures extracted using the chiral condensate as the order parameter, while the solid circles correspond to those obtained from the chiral susceptibility. All temperatures are shifted by subtracting the pseudo-critical temperature $T_{c} $ in the chiral limit.}
  \label{fig:σ-Xm}
\end{figure}

Similar calculations can be imposed for the other values of the quark mass. To show the tendency of the chiral susceptibility, we give the results for various pion masses from Model III in Fig. \ref{fig:XM-0-150}. From the figure, we can see that when the quark mass increases, the height of the peaks of $\chi_\sigma$ decreases, while the pseudo-critical temperature (the peak location) increases. We also present the results of the pseudo-critical temperatures for all three models in Fig. \ref{fig:σ-Xm}. We observe that the critical temperatures obtained from the two different order parameters differ significantly at high temperatures, with the one from the chiral susceptibility being much more sensitive to the pion mass. Generally, the transition temperatures from the chiral susceptibility are higher than those from $d\sigma/dT$ with the same quark mass. It is self-consistently that the differences of the two transition temperatures vanish towards the chiral limit, when the transition becomes a rigorous one and an exact transition temperature can be defined. For the effect of the model setting, it is easy to see that both the transition temperatures from the two order parameters in Model III are more sensitive to the pion mass than those in Model I and II. In fact, this is why we adopt the model proposed in this work to study the dependence of the critical temperature, extracted from the chiral susceptibility, on both the quark mass and the pion mass. Since the chiral susceptibility is widely used in both functional renormalization group calculations and lattice QCD,  we will identify the pseudo-critical temperature $T_c$ with the maximum($\chi_{M}^{max}$) of the chiral susceptibility in later calculations, in order to give a straightforward comparison with the other methods.

\subsection{ $T_c$ Scaling}

Generally, with a small value of quark mass, the physics near the transition temperature is still governed by the critical scaling behavior. Thus, for small quark masses, the corresponding pseudo-critical temperature also obeys a scaling law as
\begin{equation}
T_c=\alpha \times m_q^{1/\Delta}+T_{c0} ,
\label{eq:Tc-mq}
\end{equation}
with $ \Delta = {\beta\delta}$ , $T_{c0}$ the critical temperature in the chiral limit and $\alpha$ a coefficient depending on the models. This scaling law can be derived from the near critical scaling of $\sigma$. Since the holographic model gives the mean-field critical exponents $\beta=1/2,\delta=3$, we have $\Delta=3/2$. As proved in \cite{Erlich:2005qh}, the Gell-Mann–Oakes–Renner (GOR) relation $m_{\pi}^{2} f_{\pi}^{2}=2 m_{q} \langle \bar{q} q \rangle$ is satisfied in the soft-wall model, thus we have $m_\pi^2\propto m_q$ and $T_c=m_\pi^{4/3}+T_{c0}$.

\begin{figure}[htb]
  \centering

  \begin{subfigure}{\linewidth}
    \includegraphics[height=0.20\textheight]{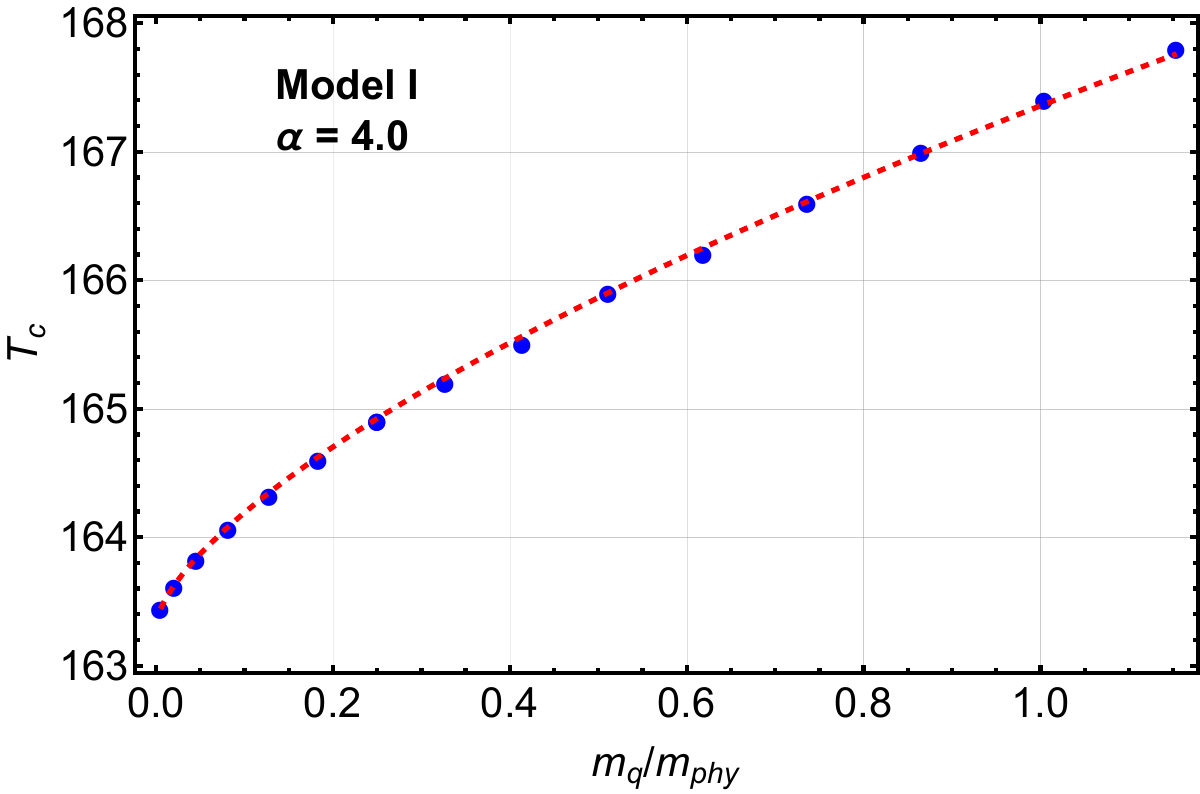}

    \label{fig:Tc-mq-1}
  \end{subfigure}
  \vskip 1em

  \begin{subfigure}{\linewidth}
    \includegraphics[height=0.20\textheight]{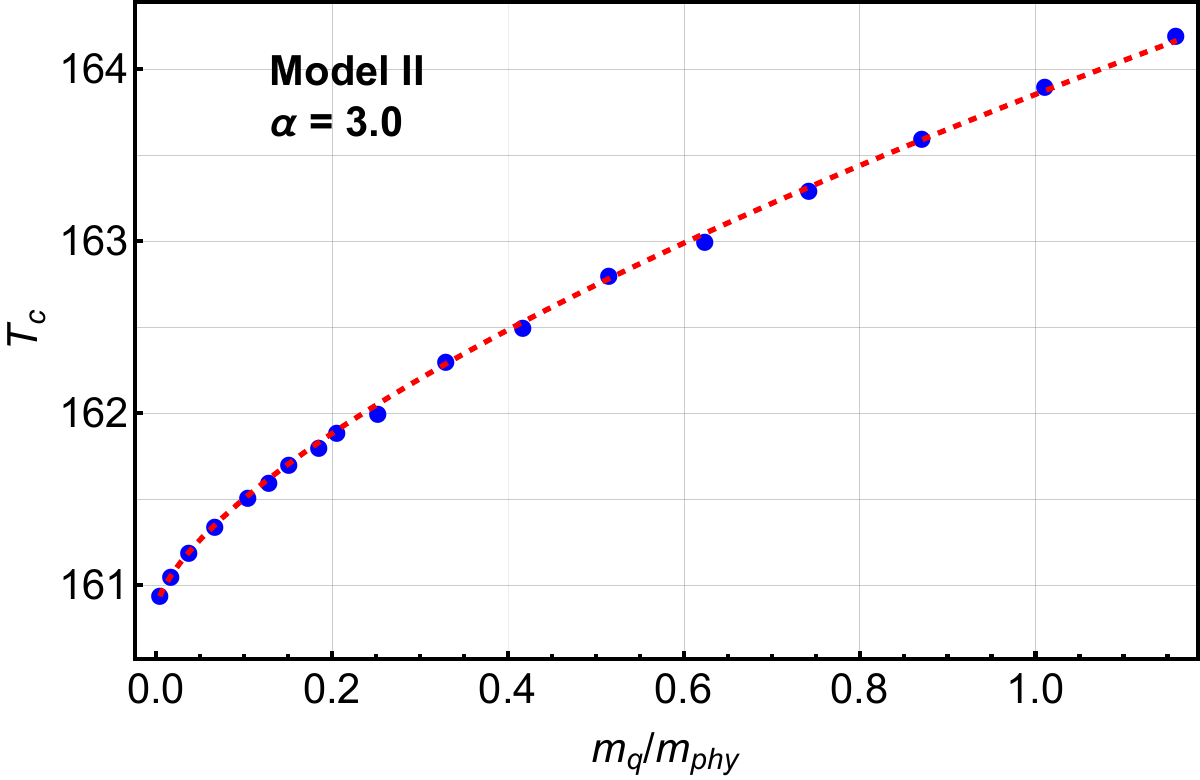}

    \label{fig:Tc-mq-2}
  \end{subfigure}
  \vskip 1em

  \begin{subfigure}{\linewidth}
    \includegraphics[height=0.20\textheight]{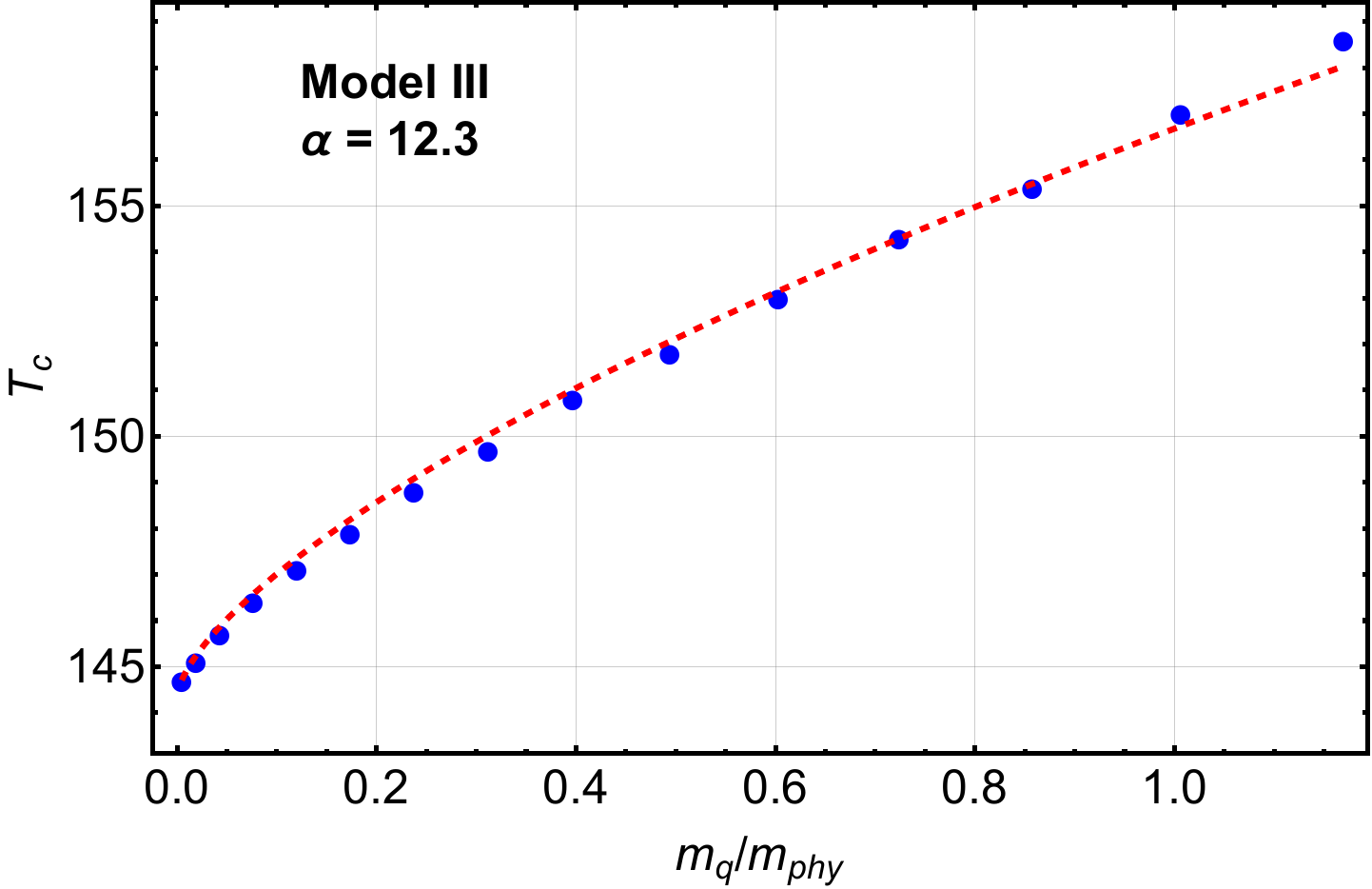}

    \label{fig:Tc-mq-3}
  \end{subfigure}

  \caption{The dependence of the pseudo-critical temperature  $T_c$  on the scaled quark mass $m_q/m_{phy}$, with the physical quark mass $m_{phy}$ corresponding to $m_{\pi} = 139.6~\mathrm{MeV} $. The red dashed curve represents the fit obtained using the data points.
}
  \label{fig:Tc-mq}
\end{figure}

As an explicit check of this scaling behavior, we extract the pseudo-critical temperatures as functions of quark mass for the three models, as shown in the blue dots in Fig.\ref{fig:Tc-mq}. The upper, middle, and lower panels show the results in Model I, II, and III, respectively. There, we show that the pseudo-critical temperatures can be well described by the scaling law (the red dashed lines) with $\alpha=4.07$ for Model I, $\alpha=3.03$ for Model II, and $\alpha=11.73$ for Model III, with $T_{c0}$ taking the values in the chiral limit of the corresponding models.  From the fitting, it is interesting to see that the `slope' $\alpha$ in Model III is much larger than that in Model I and Model II.

\begin{figure}[htb]
  \centering
  \includegraphics[width=0.48\textwidth]{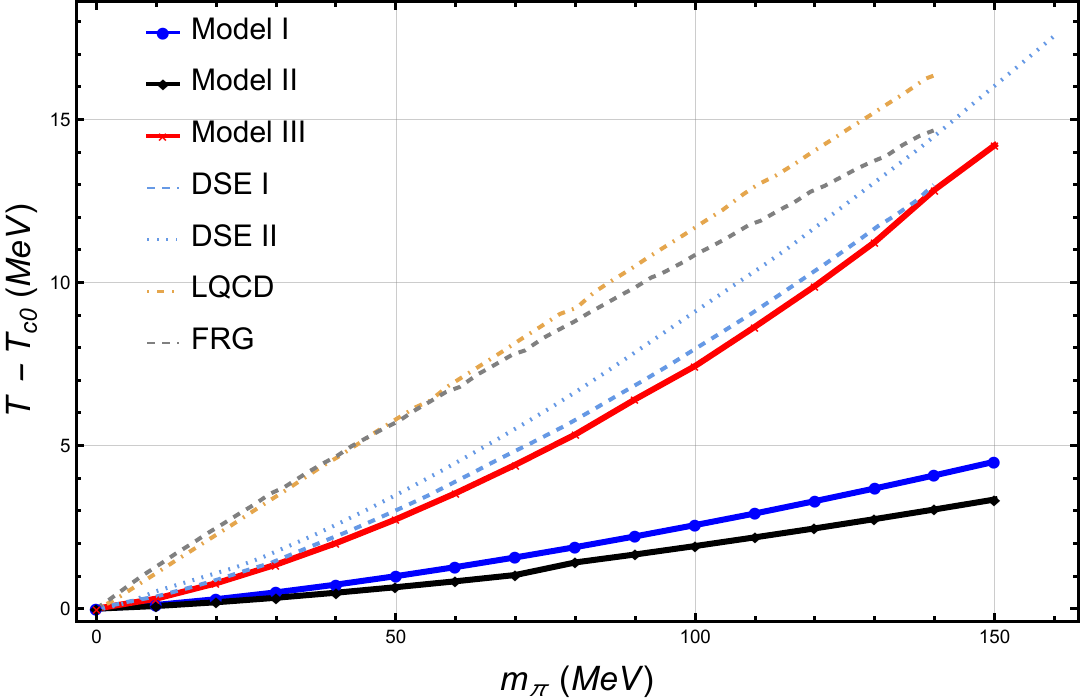}
  \caption{The pseudo-critical temperature as a function of $m_\pi$. The lattice QCD results are taken from Ref. ~\cite{HotQCD:2019xnw}, the functional renormalization group (fRG) results are taken from Ref. ~\cite{Braun:2020ada}, and the Dyson-Schwinger equations (DSE) approach results are taken from Ref. ~\cite{Bai:2020ufa}.
}
  \label{fig:DSE-FRG-LQCD}
\end{figure}

In Fig.~\ref{fig:DSE-FRG-LQCD}, we present the results for the pseudo-critical temperature $T_c$ as a function of pion mass. Due to the limited availability of \( N_f = 2 \) data, we take $N_f = 2+1$ results from lattice QCD (the orange-brown dot-dashed line, from \cite{HotQCD:2019xnw}), FRG (the medium gray dashed line, from ~\cite{Braun:2020ada}), and DSE approaches (the light blue dashed line and light blue dotted line, from ~\cite{Bai:2020ufa}) for comparison. We have subtracted the $T_c$ data from $T_{c0}$ in order to get rid of the effect of differences in $T_{c0}$.  From the figure, it can be seen that, unlike LQCD and FRG results, our model does not exhibit a linear dependence. The linear dependence observed in the FRG and LQCD results is because their critical exponents are beyond the mean-field level of 3D $O(4)$ universal class. But it is consistent with the conclusion obtained from the DSE using mean-field exponents. 

Besides the match of scaling exponent $\Delta$ with the DSE results for all three models, it is obvious that the increasing rates in Model I and Model II are too small compared with the DSE results, as well as the lattice simulations. However, by doing a further modification to the scalar potentials, we find a way to cure this problem in Model III. The logarithmic term and the $r$-dependent coupling in the scalar potential soften the potential, thereby broadening the continuous transition and making the scaling slope steeper. This is why we try to propose a new modification to the soft-wall model. We note that the scaling of $T_c$ should be taken into account in model construction within the soft-wall framework.

\begin{figure}[htb]
  \centering

  \begin{subfigure}{\linewidth}
    \includegraphics[height=0.20\textheight]{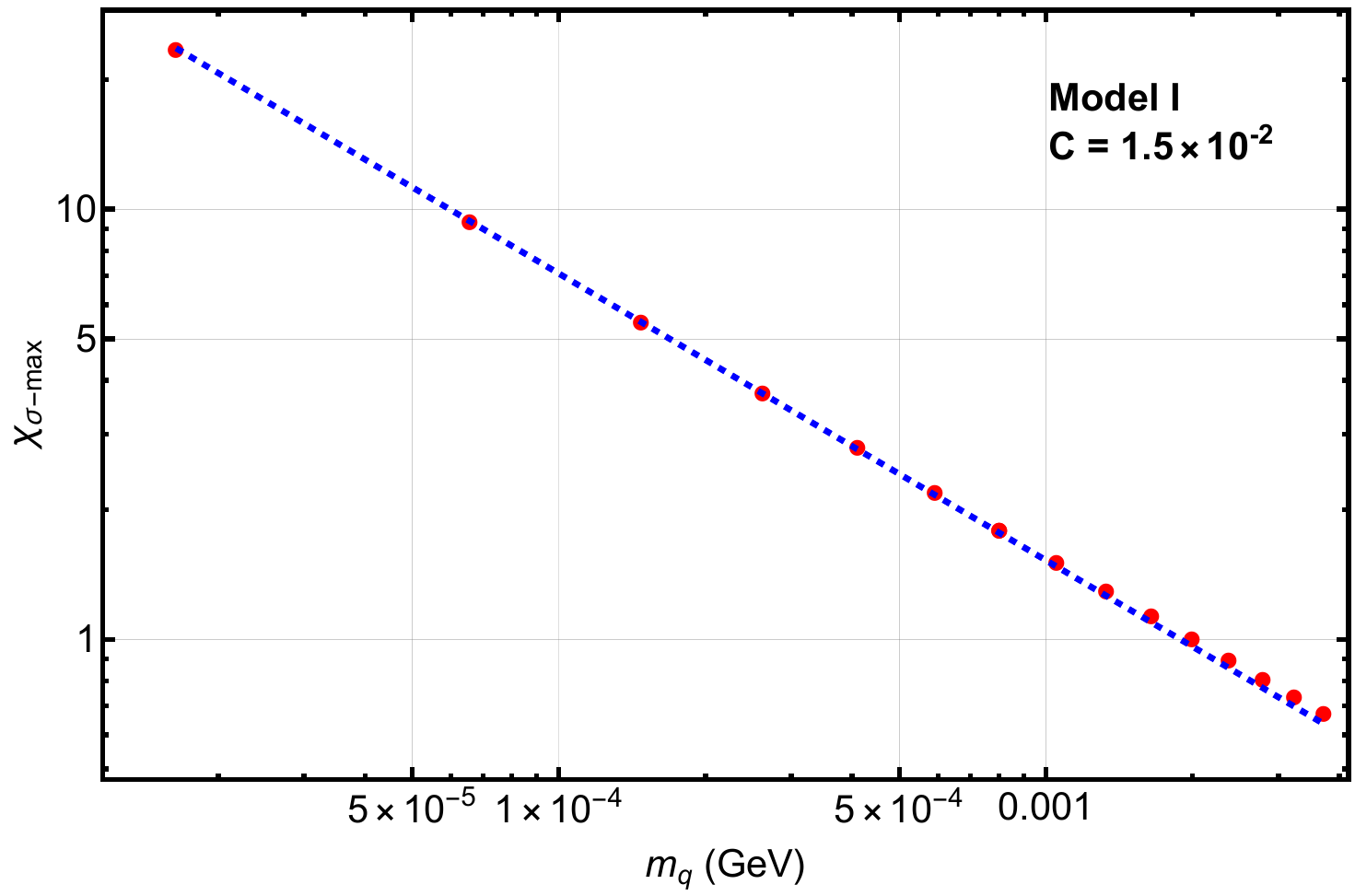}

    \label{fig:Xm-max-mq-1}
  \end{subfigure}
  \vskip 1em

  \begin{subfigure}{\linewidth}
    \includegraphics[height=0.20\textheight]{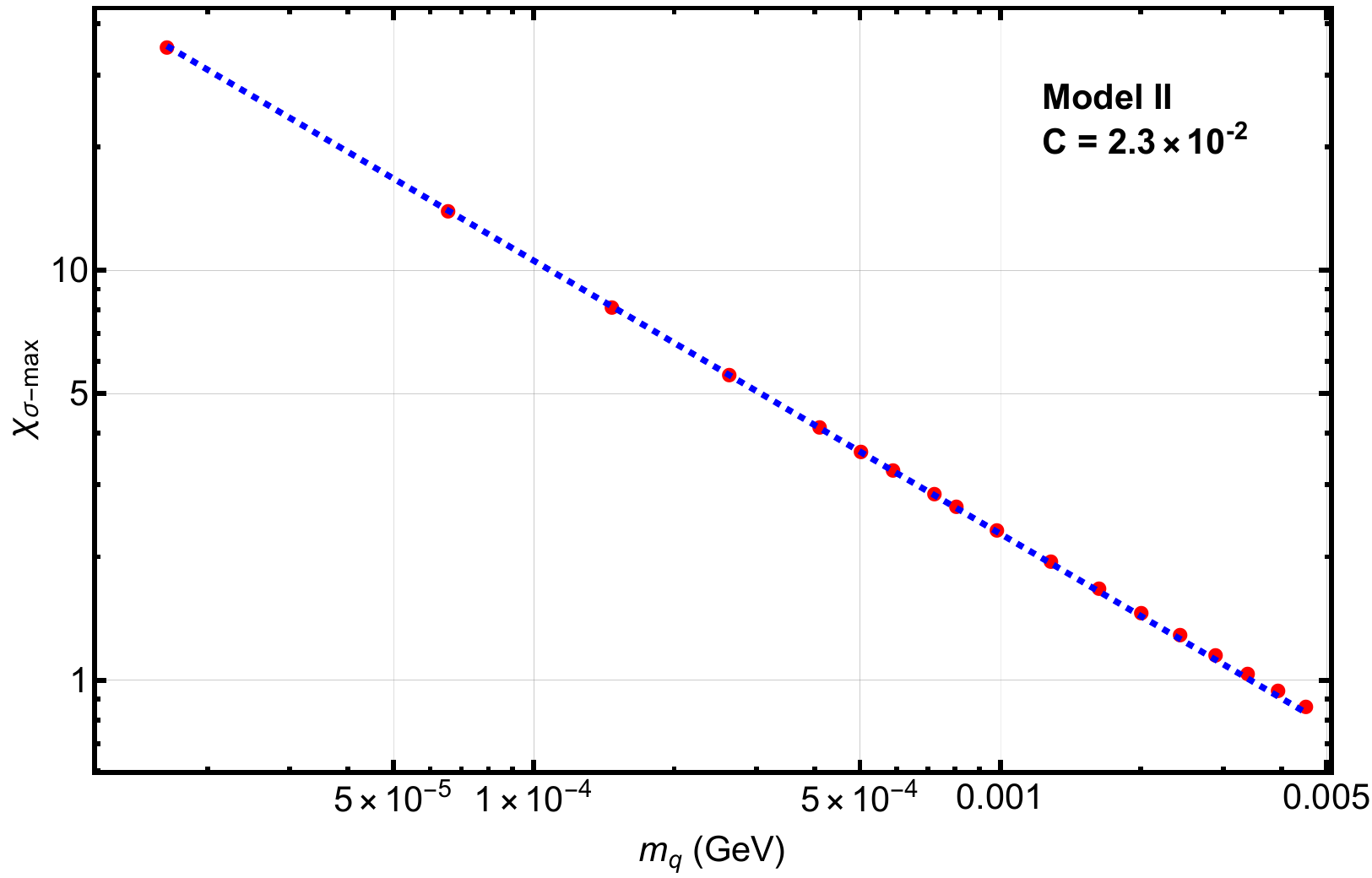}

    \label{fig:Xm-max-mq-2}
  \end{subfigure}
  \vskip 1em

  \begin{subfigure}{\linewidth}
    \includegraphics[height=0.20\textheight]{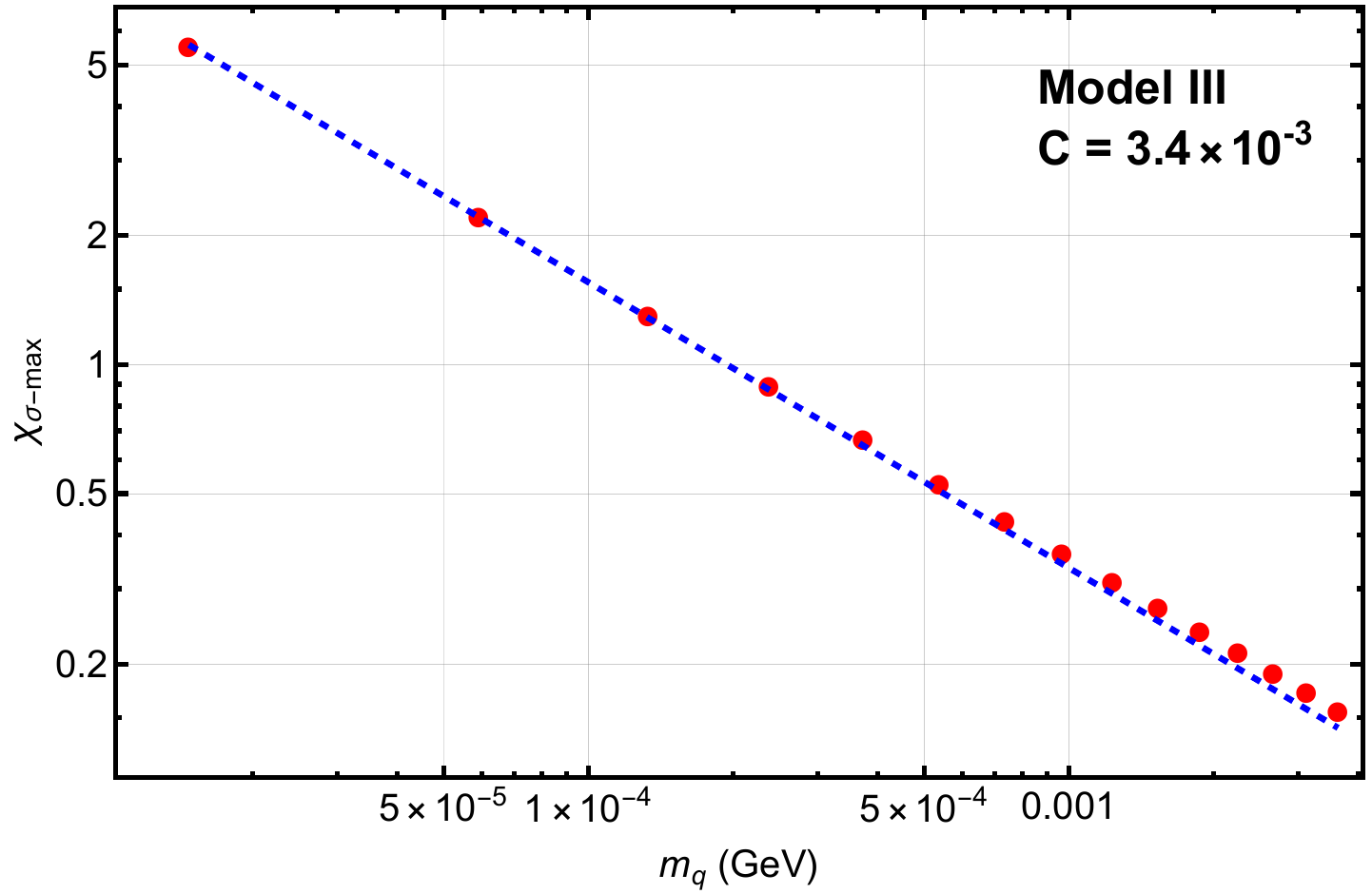}

    \label{fig:Xm-max-mq-3}
  \end{subfigure}

  \caption{Dependence of the maximum chiral susceptibility $\chi_{\sigma,\text{max}}$ on the quark mass.
}
  \label{fig:Xm-max-mq}
\end{figure}

Furthermore, in Fig.~\ref{fig:Xm-max-mq} we show the relation between the maximum chiral susceptibility $\chi_{\sigma,max}$ and $ m_q $. According to Ref.~\cite{Holl:1998qs}, they obey the following scaling behavior

\begin{equation}
\chi_{\sigma,max}=C \times m_q^{1/ \delta-1} .
\label{eq:Xm-mq}
\end{equation}
It can be seen from the figure that the scaling law (the blue dotted lines) fits the numerical data quite well. The coefficient $C$ is fitted as $C=0.016$ for Model I, $C=0.023$ for Model II, and $C=0.0034$ for Model III. Again, we see that the coefficient in Model III has a different order from Model I and Model II, which might be important in improving the $T_c$ scaling with pion mass. 

Therefore, we have seen that all the scaling laws are satisfied in the soft-wall AdS/QCD models, though all the exponents are at the mean-field level. It can also be seen that the models with a simple power of the scalar potential cannot well describe the coefficients (or the `slopes'). We propose a way to improve this behavior and get comparable $T_c$ scaling in both the exponents and the coefficients, with those obtained from the other non-perturbative methods.

\subsection{The transition temperature towards chiral limit}

Before closing this section, we will discuss another interesting topic. As we know, it is more expensive for lattice simulations to work with a smaller quark mass. In particular, it is very hard in the chiral limit. Thus, different ways have been developed to extrapolate the finite quark mass calculation to get reliable results in the chiral limit. To get the critical temperature in the chiral limit at a lower cost, it is better to find certain observables from which the critical temperatures converge to that in the chiral limit more rapidly. As shown in lattice simulations \cite{HotQCD:2019xnw}, the ratio $R=f_\chi(z=0)/f_{\chi,\text{max}} $ coincides with the ratio of the chiral susceptibility evaluated at the critical temperature in the chiral limit to its peak value, i.e. $\chi_{\sigma}(T=T_{c0})/\chi_{\sigma,\text{max}}=f_\chi(z=0)/f_{\chi,\text{max}}=R $ for small quark masses. Such a conclusion comes from the fact that the pseudo-critical temperatures for small quark masses are governed by the critical point and thus controlled by $f_\chi$ as well ( $z=0$ characterizes the critical point ). A direct consequence of this relation is that $T_{c0}$ can be obtained from calculations at finite quark masses from the condition $\chi_{\sigma}(T=T_{c0})=R \chi_{\sigma, max}$ (here, we note that there are two temperatures satisfying this condition, and one has to choose the lower one). In this section, we provide explicit holographic confirmation of this relation.

\begin{figure}[htb]
  \centering
  \includegraphics[width=0.48\textwidth]{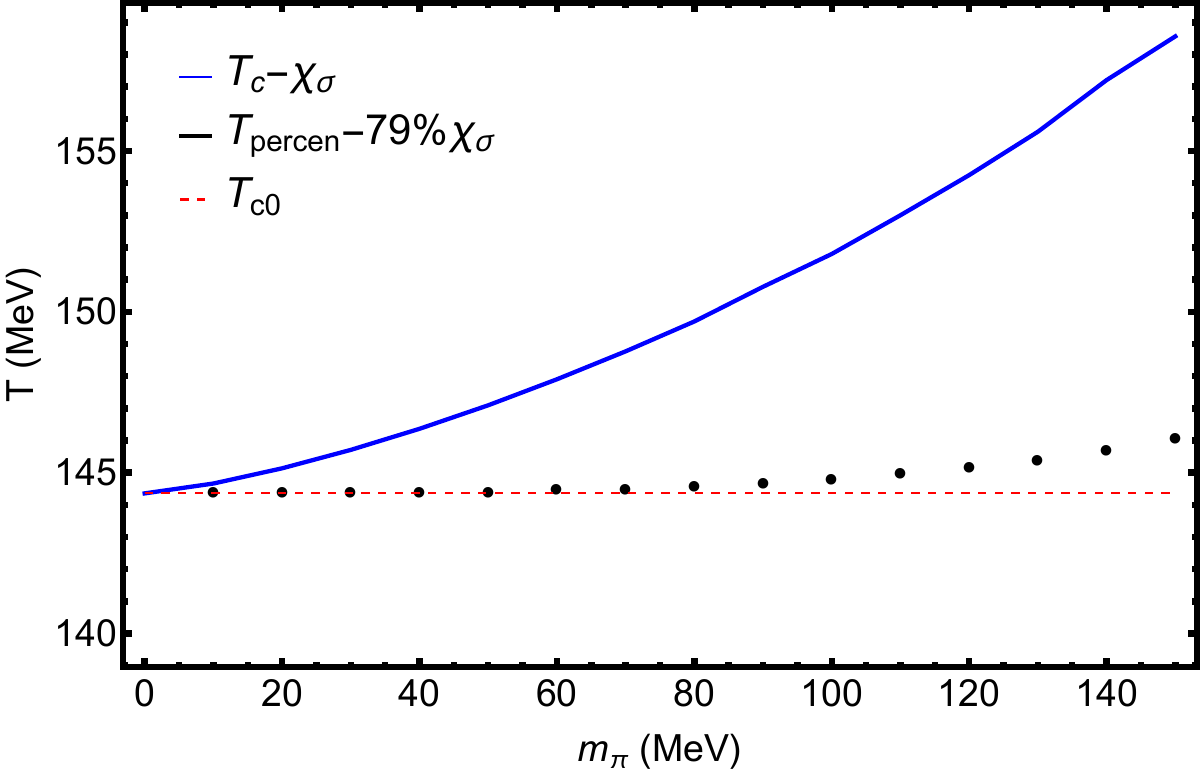}
  \caption{The relation between the pion mass \( m_\pi \) and the temperature at which \( \chi_\sigma \) reaches 79\% of its maximum value.
  }
  \label{fig:T-percent}
\end{figure}

From the previous discussion, we have $R=f_\chi(0)/f_{\chi,max}\simeq 79\%$ in the soft-wall model. Thus, $T_{c0}$ is expected to correspond to the $79\%$ height of the peaks at finite quark masses. As an explicit check,  in Fig.~\ref{fig:T-percent}, we compare the temperatures corresponding to the peak of $\chi_\sigma $ (the blue solid line) for different pion masses with those obtained from the $79\%$ height of the peak (the black dots), in reference to our critical temperature  $T_{c0}$ (the red dashed line). The figure clearly shows that the percentage-temperature $T_{\text{percent}}$  approaches $T_{c0}$ much faster than $T_c$. For $m_\pi<60\rm{MeV}$, $T_{\text{percent}}$ almost reaches $T_{c0}$, while $T_c=T_{c0}$ only in the chiral limit $m_\pi=0$. Here, although we have so far presented results only for Model III, we have checked that the relation holds in Models I and II as well. Therefore, the holographic calculation confirms that the percentage-temperature is a good observable for extracting $T_{c0}$ from finite quark mass calculations.

\section{ CONCLUSION AND DISCUSSION}\label{CONCLUSION AND DISCUSSION}

The critical point of the two-flavor soft-wall AdS/QCD model is systematically studied in Ref.~\cite{Chen:2018msc}, and it is proved to exhibit mean-field scaling laws with exponents $\beta=1/2, \delta=3$, when tuning either the temperature or the quark mass in its vicinity. However, the universal scaling functions that arise when temperature and quark mass are varied simultaneously, together with the scaling behavior of the pseudo-critical temperature itself, remain largely unexplored in this model. To verify the effectiveness and internal consistency of this holographic model, we present a careful study of the near-critical scaling behavior of various chiral order parameters. 

Under different model settings and physical conditions, we find that the rescaled magnetization $M/h^{1/\delta}$ in the chiral limit collapses onto a single, universal curve in the chiral limit $m_q\rightarrow0$, defining the scaling function $f_G(z)$. Quantitatively, this scaling function obtained numerically from holography is in excellent agreement with its four-dimensional mean-field counterpart. Furthermore, we derive a perturbative equation to evaluate the chiral susceptibility, from which we obtain another universal scaling function $f_\chi(z)$.  It is found that the universal relationship $ f_\chi(z)=\frac1\delta {f_G(z)} -\frac z{\beta\delta} f_G^{\prime}(z)$ holds in all three models, confirming the theoretical consistency of this holographic framework.

Furthermore, we extract the pseudo-critical temperature from the peak of the susceptibility and from the inflection point of the chiral condensate.
In contrast to the claim in Ref.~\cite{Gao:2016qkh}, the three models yield markedly different pseudo-critical temperatures when extracted from the two order parameters, although the results coincide in the chiral limit. In this context, our results are consistent with the view in Ref.~\cite{Aarts:2020vyb}, which suggests that different order parameters can yield significantly different pseudo-critical temperatures. Adopting the susceptibility-peak definition of the transition temperature, we numerically verify that in all three holographic models the pseudo-critical temperature obeys the universal scaling law $T_c-T_{c0}\sim m_q^{1/\beta\delta}$ or $T_c-T_{c0}\sim m_\pi^{2/\beta\delta}$. The scaling region in these models can extend below the physical quark mass. These findings support the consistency of soft-wall models with mean-field universality. Moreover, we find that the percentage-temperature defined by the $79 \%$ (the value is set by the ratio $f_\chi(0)/f_\chi(T_c)$ level of the chiral-susceptibility peak), converges to its chiral-limit value much faster than the pseudo-critical temperature, offering a practical observable for determining $T_{c0}$.

While the three holographic models share identical critical exponents and scaling functions in the chiral limit, their scaling coefficients span orders of magnitude. Compared with $T_c$ scaling from DSE~\cite{Bai:2020ufa}, FRG~\cite{Braun:2020ada}, and lattice QCD~\cite{HotQCD:2019xnw}, Models I and II yield a slope—i.e., the scaling coefficient—that is roughly one-third as large, a discrepancy that appears to afflict earlier soft-wall constructions as well. It is interesting to see that by softening the scalar potential with certain logarithmic terms and $r$-dependent coefficients, Model III reproduces a $T_c$ scaling in quantitative agreement with the DSE results \cite{Bai:2020ufa}. These findings might offer valuable guidance for refining future model constructions in the soft-wall AdS/QCD. In this study, we have restricted our attention to the two-flavour chiral critical point. Extending the analysis to the entire critical lines in the quark-mass phase diagram, and examining the resulting consequences for the critical end point within the same holographic framework, constitutes an interesting direction for future work.

\vspace*{1cm}

{\bf Acknowledgements}\quad
This work is supported by the National Natural Science Foundation of China under Grant Nos. 12275108, 12235016.

\vspace*{1cm}

\bibliographystyle{apsrev4-1}
\bibliography{refs}
\end{document}